\documentclass[%
 reprint,
 superscriptaddress,
 preprintnumbers,
 bibnotes,
 amsmath,amssymb,
 aps,
 prl,
]{revtex4-1}

\usepackage[utf8]{inputenc}
\usepackage{graphicx}
\usepackage{tikz}
\usetikzlibrary{snakes}
\usetikzlibrary{calc}
\usetikzlibrary{decorations}
\usepackage{dcolumn}
\usepackage{bm}
\usepackage{hyperref}
\hypersetup{
    colorlinks=true,
    linkcolor=black,
    filecolor=black,      
    urlcolor=black,
    citecolor=black,
}

\newcommand{\llangle}{\langle\!\langle}
\newcommand{\rrangle}{\rangle\!\rangle}

\begin{document}


\title{Feynman Integrals and Scattering Amplitudes from Wilson Loops}
\author{Song He}
\email{songhe@itp.ac.cn}
\affiliation{%
CAS Key Laboratory of Theoretical Physics, Institute of Theoretical Physics, Chinese Academy of Sciences, Beijing 100190, China
}%
\affiliation{
School of Fundamental Physics and Mathematical Sciences, Hangzhou Institute for Advanced Study, UCAS, Hangzhou 310024, China}
\affiliation{ICTP-AP
International Centre for Theoretical Physics Asia-Pacific, Beijing/Hangzhou, China}
\affiliation{%
School of Physical Sciences, University of Chinese Academy of Sciences, No.19A Yuquan Road, Beijing 100049, China
}%
\author{Zhenjie Li}
 \email{lizhenjie@itp.ac.cn}
 \affiliation{%
CAS Key Laboratory of Theoretical Physics, Institute of Theoretical Physics, Chinese Academy of Sciences, Beijing 100190, China
}%

\affiliation{%
School of Physical Sciences, University of Chinese Academy of Sciences, No.19A Yuquan Road, Beijing 100049, China
}
 \author{Qinglin Yang}%
 \email{yangqinglin@itp.ac.cn}
 \affiliation{%
CAS Key Laboratory of Theoretical Physics, Institute of Theoretical Physics, Chinese Academy of Sciences, Beijing 100190, China
}%
\affiliation{%
School of Physical Sciences, University of Chinese Academy of Sciences, No.19A Yuquan Road, Beijing 100049, China
}%
\author{Chi Zhang}%
 \email{chi.zhang@nbi.ku.dk\\}
 \affiliation{Niels Bohr International Academy, Niels Bohr Institute, Copenhagen University, Blegdamsvej 17, 2100 Copenhagen \O{}, Denmark}

\date{\today}

\begin{abstract}
We study Feynman integrals and scattering amplitudes in ${\cal N}=4$ super-Yang-Mills theory by exploiting the duality with null polygonal Wilson loops. Certain Feynman integrals, including one-loop chiral pentagon and generic two-loop double pentagon, are given by Feynman diagrams of supersymmetric Wilson loops, where one can perform loop integrations and be left with simple integrals along edges. 
As the main application, we compute analytically for the first time the symbol of finite double pentagon integrals, which give finite part of two-loop MHV amplitudes and finite components of NMHV amplitudes to all multiplicities. We represent the double pentagon as a two-fold $\mathrm{d} \log$ integral of a one-loop hexagon, and the non-trivial part of integrations concerns rationalizing square roots contained in the latter. We obtain beautiful ``algebraic words'' which contain $6$ algebraic letters for each of the $16$ square roots, and they all nicely cancel in combinations for MHV amplitudes and NMHV components which are free of square roots. We express the final answer in terms of only two independent weight-$3$ integrable symbols, written in a remarkably compact form. In addition to algebraic letters, the alphabet consists of $164$ rational letters.
\end{abstract}

\maketitle

\section{Introduction}
Scattering amplitudes are central objects in fundamental physics: they are crucial for connecting theory to experiments in particle accelerators such as Large Hadron Collider, and they play a central role in discovering new structures of Quantum Field Theory (QFT) and gravity. As arguably the simplest QFT, tremendous progress has been made for  planar ${\cal N} = 4$ supersymmetric Yang-Mills theory (SYM); not only have hidden mathematical structures for all-loop integrands been unraveled~\cite{ArkaniHamed:2010kv, ArkaniHamed:2012nw,Arkani-Hamed:2013jha}, but the (integrated) amplitudes have also been computed to impressively high loops, for $n=6,7$~\cite{Dixon:2011pw,*Dixon:2014xca,*Dixon:2014iba,*Drummond:2014ffa,*Dixon:2015iva,*Caron-Huot:2016owq} and for higher multiplicities ~\cite{CaronHuot:2011ky,Zhang:2019vnm,He:2020vob}. Moreover, these studies have made ${\cal N}=4$ SYM an extremely fruitful playground for new methods of evaluating Feynman integrals, which is a subject of enormous interests ({\it c.f.} \cite{Bourjaily:2018aeq,Henn:2018cdp,Herrmann:2019upk} and references there in). 

In planar ${\cal N}=4$ SYM, a remarkable duality between (MHV) scattering amplitudes and null polygonal Wilson loops (WL) was discovered at both strong~\cite{Alday:2007hr,*Alday:2007he,*Alday:2009yn} and weak coupling~\cite{Brandhuber:2007yx,Drummond:2007aua,*Drummond:2007cf,*Drummond:2007bm,*Drummond:2008aq,*Bern:2008ap}; later it was established that super-amplitudes (after stripping off MHV tree prefactor) are dual to supersymmetric WL~\cite{CaronHuot:2010ek,Mason:2010yk}, and quite a lot of what we have learned about amplitudes are from this dual picture. Based on integrability~\cite{Beisert:2010jr} and operator product expansion (OPE) of WL~\cite{Alday:2010ku}, one can compute amplitudes at any value of the coupling around collinear limits~\cite{Basso:2013vsa}; the powerful $\bar{Q}$ anomaly equation~\cite{CaronHuot:2011kk} for computing multi-loop amplitudes~\cite{Zhang:2019vnm,He:2020vob}, was derived from the dual WL as well. In this letter, we exploit the dual picture in yet another context: the computation of certain Feynman integrals~\footnote{Similar ideas have been used in~\cite{CaronHuot:2011ky} which motivated our investigations; they have also been explored in~\cite{Brandhuber:2007yx,Anastasiou:2011zk} as well.}, and in turn, scattering amplitudes in ${\cal N}=4$ SYM, simplifies significantly in terms of supersymmetric WL. 

Recall that in the computation of (super-)WL, one inserts fields in the super-multiplet at edges and vertices of the null polygon, as well as chiral Lagrangians at dual points which correspond to loop variables to be integrated over~\cite{CaronHuot:2010ek}. We will see that, certain loop integrals for scattering amplitudes take particularly simple form as Feynman diagrams of WL, where it is easy to perform integrations of (some) loop insertions and left with relatively simple integrals over edge-insertions (and remaining loops). In this way, we obtain the ``$\mathrm{d}\log$'' representation for loop integrals and amplitudes made of them~\footnote{The $\mathrm{d}\log$ representation plays an important role in the study of Feynman integrals, but our WL-based form differs from those studied before~\cite{ArkaniHamed:2012nw,Herrmann:2019upk}; rather it takes a form very similar to those $\tau$ integrals in $\bar{Q}$ calculations.}, which not only makes the evaluation much simpler, but also various desired properties manifest. 
We initiate the systematic study of $\mathrm{d}\log$ representation for wide range of simple loop integrals in~\cite{He:2020uxy}, but here we focus on the computation of a class of particularly important integrals, the double pentagons~\cite{ArkaniHamed:2010gh}. We denote such an integral as $I_{\rm dp}(i,j,k,l)$ with massless corners $i,j,k,l$ (IR finite for $j>i{+}1$ and $l>k{+}1$):
\begin{align*}
I_{\mathrm{dp}}(i,j,k,l) \,\, = \,\,
\begin{tikzpicture}[baseline={([yshift=-.5ex]current bounding box.center)},scale=0.16]
\draw[black,thick](0,0)--(0,5)--(4.75,6.54)--(7.69,2.50)--(4.75,-1.54)--cycle;
\draw[black,thick](0,5)--(-4.75,6.54)--(-7.69,2.50)--(-4.75,-1.54)--(0,0);
\draw[decorate, decoration=snake, segment length=12pt,segment amplitude=2pt,black,thick] (4.75,6.54)--(4.75,-1.54);
\draw[decorate, decoration=snake, segment length=12pt,segment amplitude=2pt,black,thick] (-4.75,6.54)--(-4.75,-1.54);
\draw[black,thick](-0.9,6.5)--(0,5)--(0.9,6.5);
\filldraw[black]  (0,6) circle [radius=1.5pt];
\filldraw[black]  (-0.25,6) circle [radius=1.5pt];
\filldraw[black]  (0.25,6) circle [radius=1.5pt];
\draw[black,thick](-0.9,-1.5)--(0,0)--(0.9,-1.5);
\filldraw[black]  (0,-1) circle [radius=1.5pt];
\filldraw[black]  (-0.25,-1) circle [radius=1.5pt];
\filldraw[black]  (0.25,-1) circle [radius=1.5pt];
\draw[black,thick](4.75,6.54)--(6,7.94);
\draw[black,thick](4.75,-1.54)--(6,-3.04);
\draw[black,thick](9.19,1.6)--(7.69,2.50)--(9.19,3.4);
\draw[black,thick](-4.75,6.54)--(-6,7.94);
\draw[black,thick](-4.75,-1.54)--(-6,-3.04);
\draw[black,thick](-9.19,1.6)--(-7.69,2.50)--(-9.19,3.4);
\filldraw[black]  (9,2.5) circle [radius=1.5pt];
\filldraw[black]  (9,2.25) circle [radius=1.5pt];
\filldraw[black]  (9,2.75) circle [radius=1.5pt];
\filldraw[black]  (-9,2.5) circle [radius=1.5pt];
\filldraw[black]  (-9,2.25) circle [radius=1.5pt];
\filldraw[black]  (-9,2.75) circle [radius=1.5pt];
\filldraw[black] (6,7.94) node[anchor=south west] {{$k$}};
\filldraw[black] (6,-3.04) node[anchor=north west] {{$l$}};
\filldraw[black] (-6,7.94) node[anchor=south east] {{$j$}};
\filldraw[black] (-6,-3.04) node[anchor=north east] {{$i$}};
\end{tikzpicture}
\end{align*}
Remarkably, the two-loop MHV amplitude is given by the sum of $I_{\rm dp}(i,j,k,l)$ with $i<j<k<l<i$ cyclically (including divergent boundary terms)~\cite{ArkaniHamed:2010gh}, and these integrals alone also give a large class of components of two-loop NMHV amplitudes. While this fact can be derived from local-integral representation of amplitudes~\cite{ArkaniHamed:2010gh}, let us review pictorially how $I_{\rm dp}$ naturally give NMHV components of supersymmetric WL. Recall that the polygonal WL are most nicely formulated in terms of momentum twistors~\cite{Hodges:2009hk}, which correspond to null rays of the dual spacetime and manifest the SL$(2,2)$ dual conformal symmetries~\cite{Drummond:2006rz,*Drummond:2008vq,*Korchemsky:2010ut}: the vertices are given by $(x_{i{+}1}-x_i)^{\alpha \dot{\alpha}}=\lambda^\alpha_i \tilde\lambda^{\dot{\alpha}}_i$, 
and similarly for the Grassmann part $(\theta_{i{+}1}-\theta_i)^{\alpha I}=\lambda_i^\alpha \eta_i^I$. The (super-) momentum twistors are defined as
$$
 \mathcal{Z}_{i}=(Z_{i}^{a}\vert \chi_{i}^{A}):=(\lambda_{i}^{\alpha},x_{i}^{\alpha\dot{\alpha}}\lambda_{i\alpha}\vert\theta_{i}^{\alpha A}\lambda_{i\alpha}) \:.$$
Consider a class of components $\chi^A_i \chi^B_j\chi^C_k \chi^D_l$ for NMHV super-WL, with {\it non-adjacent} $i<j<k<l$, and it is easy to see that such a component is given by the difference of two Feynman diagrams of WL(Fig.~\ref{figwl}.). 
\begin{figure}[htbp]
\begin{center}
\begin{tikzpicture}[scale=0.5]
\filldraw[black] (-2.5,-0.25) node[anchor=east] {{$\mathcal{W}^{(2)}_{n,k=1}\biggm|_{\chi_i^A\chi_j^B\chi_k^C\chi_l^D}=$}};
\draw[black,very thick](1,1.5)--(2.25,0);
\draw[black,very thick](-1,1.5)--(-2.25,0);
\draw[black,very thick](1,-2)--(2.25,-0.5);
\draw[black,very thick](-1,-2)--(-2.25,-0.5);
\draw[blue,very thick] (1.625,0.75)--(0.71,-0.25)--(1.625,-1.25);
\draw[blue,very thick] (-1.625,0.75)--(-0.71,-0.25)--(-1.625,-1.25);
\draw[red, dashed,ultra thick](-0.71,-0.25)--(0.71,-0.25);
\filldraw[gray]  (-0.71,-0.25) circle [radius=2pt];
\filldraw[gray]  (0.71,-0.25) circle [radius=2pt];
\filldraw[black] (2.35,-0.25) node[anchor=west] {{$-$}};
\filldraw[black] (-1.625,0.75) node[anchor=south east] {{$j$}};
\filldraw[black] (-1.625,-1.25) node[anchor=north east] {{$i$}};
\filldraw[black] (1.625,0.75) node[anchor=south west] {{$k$}};
\filldraw[black] (1.625,-1.25) node[anchor=north west] {{$l$}};
\filldraw[black] (-1.48,0.68) node[anchor=west] {{$\tilde{\psi}_B$}};
\filldraw[black] (-1.48,-1.18) node[anchor=west] {{$\tilde{\psi}_A$}};
\filldraw[black] (1.48,0.68) node[anchor=east] {{$\tilde{\psi}_C$}};
\filldraw[black] (1.48,-1.18) node[anchor=east] {{$\tilde{\psi}_D$}};
\filldraw[black] (-0.73,-0.25) node[anchor=east] {{$\ell_1$}};
\filldraw[black] (0.73,-0.25) node[anchor=west] {{$\ell_2$}};
\draw[black,very thick](6.75,1.5)--(8,0);
\draw[black,very thick](4.75,1.5)--(3.5,0);
\draw[black,very thick](6.75,-2)--(8,-0.5);
\draw[black,very thick](4.75,-2)--(3.5,-0.5);
\draw[blue,very thick] (7.375,0.75)--(5.75,0.25)--(4.125,0.75);
\draw[blue,very thick] (7.375,-1.25)--(5.75,-0.75)--(4.125,-1.25);
\draw[red, dashed,ultra thick](5.75,0.25)--(5.75,-0.75);
\filldraw[gray]  (5.75,-0.75) circle [radius=2pt];
\filldraw[gray]  (5.75,0.25) circle [radius=2pt];
\filldraw[black] (4.125,0.75) node[anchor=south east] {{$j$}};
\filldraw[black] (4.125,-1.25) node[anchor=north east] {{$i$}};
\filldraw[black] (7.375,0.75) node[anchor=south west] {{$k$}};
\filldraw[black] (7.375,-1.25) node[anchor=north west] {{$l$}};
\filldraw[black] (4.25,0.75) node[anchor=north] {{$\tilde{\psi}_B$}};
\filldraw[black] (4.25,-1.25) node[anchor=south] {{$\tilde{\psi}_A$}};
\filldraw[black] (7.25,0.75) node[anchor=north] {{$\tilde{\psi}_C$}};
\filldraw[black] (7.25,-1.25) node[anchor=south] {{$\tilde{\psi}_D$}};
\filldraw[black] (5.75,-0.75) node[anchor=north] {{$\ell_1$}};
\filldraw[black] (5.75,0.25) node[anchor=south] {{$\ell_2$}};
\end{tikzpicture}
\end{center}
\caption{NMHV component of super-WL as difference of two diagrams, each equals to a double-pentagon integral.}\label{figwl}
\end{figure}

\noindent The only possible external insertions are fermions $\tilde\psi_A , \cdots, \tilde\psi_D$ at edges $i, j, k, l$, respectively, and two loop insertions are nothing but Yukawa vertices, $\psi^A \psi^B \phi_{AB} (\ell_1)$ and $\psi^C \psi^D \phi_{CD} (\ell_2)$, connected by a scalar propagator $\langle \phi_{AB} (\ell_1) \phi_{CD}(\ell_2)\rangle \sim 1/(\ell_1-\ell_2)^2$; the final result is given by this ``s-channel'' diagram, minus the ``t-channel'' one~\footnote{note that no diagram with less insertion is possible, which explains why the component vanishes at tree and one-loop level, and it is finite at two-loop level.}. To see that each diagram exactly gives a double pentagon, we refer to the argument in~\cite{CaronHuot:2010ek}: after performing the integrals over fermion insertions along the edges, we obtain all propagators and the numerators (``wavy lines'') of $I_{\rm dp}$, with loop integrations over insertion points $\ell_1, \ell_2$. Thus purely from WL picture, we see that the simplest NMHV component amplitudes at two loops are given by a difference of two WL diagrams, 
$I_{\rm dp}(i,j,k,l)-I_{\rm dp}(j,k,l,i)$. 

The double pentagon $I_{\rm dp}$ has only been evaluated for $n\leq 7$ legs~\cite{Dixon:2011nj,Henn:2018cdp}~\footnote{For two-dimensional kinematics, $n=8$ double pentagons have been evaluated analytically in~\cite{Alday:2010jz}; see~\cite{Heslop:2010kq} and \cite{Caron-Huot:2013vda} for analytic results of two-loop and three-loop amplitudes in two-dimensional kinematics.}, and starting $n=8$, it generically depends on functions of kinematics that contain irreducible square roots of Gram determinants, which we call ``algebraic letters"~\cite{Henn:2018cdp}. The most general $I_{\rm dp} (i,j,k,l)$ depends on $12$ legs, $i{-}1, i, i{+}1, \cdots, l{-}1, l, l{+}1$, which has identical kinematic space as that of the chiral octagon~\cite{ArkaniHamed:2010gh}; the generic $I_{\rm dp}$ is expected to contain $16$ square roots corresponding to $16$ four-mass box configurations of the latter, and similarly for all finite degenerations. Its analytic computation is currently beyond the reach of conventional method, {\it e.g.} using Feynman parametrization. 
The most up-to-date result is the numeric computation of $I_{\rm dp}(1,3,5,7)$ with $n=8$ at a particular kinematic point~\cite{Bourjaily:2019igt}, which suggests that although each integral contains square roots, the difference does not. This surprising observation has been confirmed by an independent $\bar{Q}$ calculation for two-loop NMHV amplitudes~\cite{Zhang:2019vnm,He:2020vob}, which shows that the above component is free of square roots for any $n$. However, the $\bar{Q}$ calculation is for the full amplitude, thus has no access to individual $I_{\rm dp}$ involving algebraic letters. In this Letter, we solve this long-standing problem by evaluating the symbol~\cite{Goncharov:2010jf,*Duhr:2011zq} of most generic $I_{\rm dp}(i,j,k,l)$ with $n\geq 12$ from WL analytically. This amounts to the first all-multiplicity computation of all finite integrals for two-loop MHV amplitudes and these special components of NMHV amplitudes.

The key 
lies in the fact that we can swap order of integrations in WL diagrams: for $I_{\rm dp}$ it is possible to perform both loop integrals and be left with $4$-fold integrals over edge-insertions, but in practice a mixture of integrations turns out to be more convenient. We apply the trick only to one of two loop integrations and evaluates the other one by the usual box expansion. 
In this way, we express $I_{\rm dp}$ as a sum of $2$-fold $\mathrm{d}\log$ integrals of some weight-$2$ functions, which turn out to be similar to ${\bar Q}$ computations~\cite{CaronHuot:2011kk,Zhang:2019vnm,He:2020vob} and the predecessor~\cite{CaronHuot:2011ky}. An important technical point is that when performing the $2$-fold integrals, one needs to``rationalize'' square roots that appear in four-mass boxes
. Among other things, we find remarkably compact ``algebraic words" of its symbol containing the $16$ square roots, where for each of them, only $4$ new algebraic letters appear compared with the corresponding four-mass box. We see nicely how the square roots cancel in the difference for NMHV components (for $n=8$, it agrees perfectly with the component amplitude computed from ${\bar Q}$ method~\cite{Zhang:2019vnm}), as well as in the cyclic sum for MHV amplitudes. Even more remarkably, the complete symbol for generic $I_{\rm dp}(i,j,k,l)$ can be expressed using only {\it two} independent weight-$3$ integrable symbols, which we will present explicitly. The alphabet contains $164$ rational letters (in addition to the algebraic ones).


\section{Warm-up example: chiral pentagon}  \label{warmup}

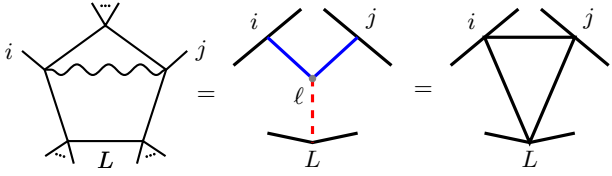
\begin{figure}[htbp]
\begin{tikzpicture}[scale=0.2]
\draw[black, thick](0,0)--(5,0)--(6.54,4.75)--(2.50,7.69)--(-1.54,4.75)--cycle;
\draw[black, thick](-1.5,-1)--(0,0)--(0.4,-1.5);
\draw[black, thick](6.5,-1)--(5,0)--(4.6,-1.5);
\draw[black, thick](6.54,4.75)--(7.94,6);
\draw[black, thick](-1.54,4.75)--(-3.04,6);
\draw[black, thick](1.6,9.19)--(2.50,7.69)--(3.4,9.19);
\draw[decorate, decoration=snake, segment length=12pt,segment amplitude=2pt,black, thick] (6.54,4.75)--(-1.54,4.75);
\filldraw[black]  (2.5,9) circle [radius=1.5pt];
\filldraw[black]  (2.25,9) circle [radius=1.5pt];
\filldraw[black]  (2.75,9) circle [radius=1.5pt];
\filldraw[black]  (-0.25,-1.03) circle [radius=1.5pt];
\filldraw[black]  (-0.5,-0.95) circle [radius=1.5pt];
\filldraw[black]  (-0.75,-0.87) circle [radius=1.5pt];
\filldraw[black]  (5.25,-1.03) circle [radius=1.5pt];
\filldraw[black]  (5.5,-0.95) circle [radius=1.5pt];
\filldraw[black]  (5.75,-0.87) circle [radius=1.5pt];
\filldraw[black] (-3.04,6) node[anchor=east] {{$i$}};
\filldraw[black] (7.94,6) node[anchor=west] {{$j$}};
\filldraw[black] (2.5,-0.1) node[anchor=north] {{$L$}};
\filldraw[black] (2.5,-0.1) node[anchor=north] {{$L$}};
\filldraw[black] (8,3) node[anchor=west] {{$=$}};
\end{tikzpicture}
\begin{tikzpicture}[scale=0.6]
\draw[black,very thick](1.5,1)--(0,2.25);
\draw[black,very thick](-2,1)--(-0.5,2.25);
\draw[black,very thick](-1.25,-0.5)--(-0.25,-0.71)--(0.75,-0.5);
\draw[blue,very thick] (0.75,1.625)--(-0.25,0.71)--(-1.25,1.625);
\draw[red, dashed,very thick](-0.25,0.71)--(-0.25,-0.71);
\filldraw[gray]  (-0.25,0.71) circle [radius=2pt];
\filldraw[black] (-1.25,1.625) node[anchor=south east] {{$i$}};
\filldraw[black] (0.75,1.625) node[anchor=south west] {{$j$}};
\filldraw[black] (-0.25,-0.71) node[anchor=north] {{$L$}};
\filldraw[black] (-0.25,0.71) node[anchor=north east] {{$\ell$}};
\filldraw[black] (1.8,0.45) node[anchor=west] {{$=$}};
\end{tikzpicture}
\begin{tikzpicture}[scale=0.6]
\draw[black,very thick](1.5,1)--(0,2.25);
\draw[black,very thick](-2,1)--(-0.5,2.25);
\draw[black,very thick](-1.25,-0.5)--(-0.25,-0.71)--(0.75,-0.5);
\draw[black,very thick](0.75,1.625)--(-0.25,-0.71)--(-1.25,1.625)--cycle;
\filldraw[black] (-1.25,1.625) node[anchor=south east] {{$i$}};
\filldraw[black] (0.75,1.625) node[anchor=south west] {{$j$}};
\filldraw[black] (-0.25,-0.71) node[anchor=north] {{$L$}};
\end{tikzpicture}
\caption{The chiral pentagon written as a WL diagram, and loop integral performed using ``star-triangle'' identity. }\label{figpenta}
\end{figure}

Before moving to $I_{\rm dp}$, let's illustrate this method using the one-loop chiral pentagon (Fig.~\ref{figpenta}), which after summing over $i<j$ gives one-loop MHV amplitude (including IR divergent terms with $j=i{+}1$). It is convenient to introduce the shorthand notation $\llangle\ell i\rrangle:=\langle \ell i{-}1 i \rangle \langle \ell i i{+}1 \rangle$ 
(loop momentum $\ell$ is represented by a line/bi-twistor). The integral has four propagators associated with $i,j$, and the last one specified by a generic line $L$:
\begin{equation}
I_{\mathrm{p}}(i,j,L):=\int \frac{\mathrm{d}^4 \ell \langle \ell\, \bar{i}\cap \bar{j}\rangle \langle L i j\rangle}{\llangle\ell i\rrangle\llangle\ell j\rrangle \langle\ell L\rangle}\,,
\end{equation}
where the numerator depends on the two solutions of the two-mass-easy Schubert problem: $(i j)$ and the intersection of planes $\bar{i}:=(i{-}1\, i\, i{+}1)$ and $\bar{j}$.  

In~\cite{CaronHuot:2010ek}, $I_{\mathrm{p}}$ was interpreted as a (bosonic) WL diagram with gluons inserted at edge $i,j$ and a Lagrangian insertion at $\ell$
.  For our purpose, it is convenient to represent $I_{\mathrm{p}}/\langle L i j\rangle$ as a WL diagram with two fermions inserted at edge $i,j$, both connected to the Yukawa vertex $\psi \psi \phi(\ell)$, and a scalar propagator from $\ell$ to the reference line $L$ (Fig~\ref{figpenta}). To see this, we write $\llangle\ell i\rrangle$ as 1-d integral $\frac{1}{\llangle\ell i\rrangle}=\int_0^\infty \frac{\mathrm{d} \tau}{\langle \ell i X(\tau)\rangle^2}$, where we have introduced the twistor interpolating between $Z_{i{-}1}$ and $Z_{i{+}1}$~\cite{CaronHuot:2010ek}: $X(\tau):=Z_{i{-}1} + \tau Z_{i{+}1}$. Note $x:=i X(\tau)$ corresponds to the insertion point on edge $i$ (with two endpoints given by $\tau\to 0, \infty$) and similarly for $y:= j Y(\tau')$; the numerator is obtained by taking into account that for fermion propagator $[i | (x-\ell) (\ell-y) | j ]  \propto \langle \ell~\bar{i} \cap \bar{j}\rangle$. Remarkably, one can easily perform the loop integration for this diagram
\begin{equation}\label{star-triangle}
\int \frac{\mathrm{d}^4 \ell \:\:\langle\ell \bar{i} \cap \bar{j}\rangle}{\langle\ell i X\rangle^2 \langle\ell j Y\rangle^2 \langle\ell L \rangle}=\frac{\langle L\bar{i} \cap \bar{j}\rangle}{\langle L i X\rangle \langle L j Y\rangle \langle i X j Y\rangle}\,,
\end{equation}
where we have used a version of ``star-triangle'' identity (Fig.~\ref{figpenta}) for three-point correlators in CFT~\cite{Chicherin:2012yn}. 
By \eqref{star-triangle} one can represent $I_{\mathrm{p}}(i,j,L)$ as a two-fold line integral
\begin{align}\label{pentagonWL}
I_{\mathrm{p}}(i,j,L)&=\int_0^\infty \mathrm{d}\tau'\int_0^\infty \mathrm{d} \tau \frac{\langle L \bar{i} \cap \bar{j}\rangle \langle L ij \rangle}{\langle L i X \rangle \langle L j Y\rangle \langle i X j Y\rangle}\nonumber\\
&=\int \mathrm{d}\log \frac{\langle L j Y\rangle }{\langle \bar i (j Y) \cap (i L)\rangle} \mathrm{d}\log \frac{\langle i X j Y\rangle}{\langle L i X\rangle} 
\end{align}
where in the second equality it is written as a integral over two $\mathrm{d}\log$'s by introducing a spurious pole, which makes it clear that it is a pure function (the integration domain for $X,Y$ are edge $i$ and $j$). 
In this form, the integrations can be performed trivially yielding the well-known result; 
this WL representation not only trivializes the evaluation of integrals, but also manifests properties of the answer, such as dual conformal invariance (DCI) and uniform transcendental weights.

\section{Double pentagon as two-fold integral} 
\label{Dp2int}

Now we turn to the main object of interests: the double pentagon integral, 
\begin{equation}\label{double-pentagon}
I_{\rm dp}(i,j,k,l):=\int \frac{\mathrm{d}^4 \ell_1 \mathrm{d}^4 \ell_2~\langle \ell_1 \bar{i}\cap \bar{j}\rangle \langle \ell_2 \bar{k}\cap \bar{l}\rangle \langle i j k l\rangle}{\llangle\ell_1 i\rrangle\llangle\ell_1 j\rrangle \langle \ell_1 \ell_2 \rangle \llangle\ell_2 k\rrangle \llangle\ell_2 l\rrangle}\,.
\end{equation}
Without explicitly using the WL diagram, we apply the same manipulation as above for loop $\ell_1$ to write it as an integration of a one-loop hexagon over edge $i,j$:
\begin{equation} \label{dpWL}
I_{\text{dp}}(i,j,k,l)=\int\frac{\mathrm{d}^2\tau \langle ijkl\rangle}{\langle iXjY\rangle}
\begin{tikzpicture}[baseline={([yshift=-.5ex]current bounding box.center)},scale=0.45]
\draw[black, thick](0,0)--(1,1.5)--(3,1.5)--(4,0)--(3,-1.5)--(1,-1.5)--cycle;
\draw[decorate, decoration=snake, segment length=10pt,segment amplitude=1.5pt,black, thick] (3,1.5)--(3,-1.5);
\draw[decorate, decoration=snake, segment length=10pt,segment amplitude=1.5pt,black, thick] (1,1.5)--(1,-1.5);
\draw[black, thick](0.4,1.9)--(1,1.5)--(1,2.3);
\draw[black, thick](0.3,-2)--(1,-1.5)--(1,-2.4);
\draw[black, thick](-0.6,0.4)--(0,0)--(-0.6,-0.4);
\draw[black, thick](3,1.5)--(3.4,2.2);
\draw[black, thick](3,-1.5)--(3.5,-2.2);
\draw[black, thick](4.6,0.4)--(4,0)--(4.6,-0.4);
\filldraw[black] (1,2.3) node[anchor=south] {{$k{-}1$}};
\filldraw[black] (1,-2.4) node[anchor=north] {{$l{+}1$}};
\filldraw[black] (0.4,1.9) node[anchor=south east] {{$j$}};
\filldraw[black] (0.3,-2) node[anchor=north east] {{$i$}};
\filldraw[black] (3.4,2.2) node[anchor=south west] {{$k$}};
\filldraw[black] (3.5,-2.2) node[anchor=north west] {{$l$}};
\filldraw[black] (4.6,0.4) node[anchor=west] {{$k$+1}};
\filldraw[black] (4.6,-0.4) node[anchor=west] {{$l$-1}};
\filldraw[black] (-0.6,0.4) node[anchor=east] {{$Y$}};
\filldraw[black] (-0.6,-0.4) node[anchor=east] {{$X$}};
\filldraw[black] (0.7,-0.6) node[anchor=north east] {{$x$}};
\filldraw[black] (0.7,0.5) node[anchor=south east] {{$y$}};
\filldraw[black] (2.1,1.5) node[anchor=south] {{$x_k$}};
\filldraw[black] (2.1,-1.4) node[anchor=north] {{$x_{l+1}$}};
\filldraw[black] (3.4,0.6) node[anchor=south west] {{$x_{k+1}$}};
\filldraw[black] (3.4,-0.6) node[anchor=north west] {{$x_l$}};
\end{tikzpicture}
\end{equation}
where the one-loop hexagon is defined as
\begin{equation}\label{hexagon}
I_{\rm hex}:=\int \frac{\mathrm{d}^4 \ell_2 \: \langle \ell_2 {\bar i} \cap {\bar j}\rangle \langle \ell_2 \bar{k} \cap \bar{l}\rangle}{\langle \ell_2 i X\rangle \langle \ell_2 j Y\rangle \llangle\ell_2 k\rrangle \llangle\ell_2 l\rrangle}\,.
\end{equation}
Note that it has two ``deformed'' legs $X, Y$ rather than original $i{+}1$ and $j{-}1$. 
The computation of $I_{\mathrm{hex}}$ is standard: using the familiar box expansion \cite{Bern:1992em,Bern:1993kr} or the general algorithm provided in \cite{Arkani-Hamed:2017ahv}. Either way, the result turn out to be a linear combination of $\binom{6}{4}=15$ box functions with some $\mathrm{d} \log$ 2-forms as the coefficients. 

To describe our result, it is convenient to label the $6$ propagators of $I_{\rm hex}$ by the $6$ points $x, y, x_k
, x_{k{+}1}
, x_{l}
, x_{l{+}1}
$ in the dual spacetime (eq.~\eqref{dpWL}) and introduce the ``$\gamma$''-deformed four-mass box function: $\tilde F:=\gamma F(u, v)-\frac 1 2 \log u \log v$, where we have introduced $\gamma:=\tfrac{r_{1}-r_{2}}{r_{1}+r_{2}}$ and
\begin{gather*}
  F(u,v):={\rm Li}_2(1-
  z)-{\rm Li_2}(1-\bar{z})+\tfrac{1}{2} \log \biggl(\frac {z}{\bar{z}}\biggr) \log (v) \\
  u=u_{a,b,c,d}=z\bar{z}\:,\:v=u_{b,c,d,a}=(1-z)(1-\bar{z})
\end{gather*}
with $u_{a,b,c,d}:=x_{a,b}^2 x_{c,d}^2/(x_{a,c}^2 x_{b,d}^2)$ and two corresponding leading singularities $r_{1},r_{2}$~\cite{ArkaniHamed:2010gh} evaluated at two solutions of the four-mass Schubert problem.
The upshot is that $I_{\rm dp}$ can be densely expressed as 
\begin{equation}\label{dp_2dint}
  \int \left([x, x_k] I_{x, x_k} {-} (k{-}1 \leftrightarrow k{+}1) \right) -(\bar{k}\leftrightarrow \bar{l}) + [x,y] I_{x,y}
\end{equation}
where the second term is obtained by swapping $k{-}1$ with $k{+}1$, the next two terms are given by swapping $k$, $x_k$ or $x_{k{+}1}$ with $l$, $x_l$ or $x_{l{+}1}$~\footnote{Throughout the letter, any relabelling acts on external legs/momentum twistors, rather than on dual points. We denote the $15$ residues of the hexagon ($2$ forms) by the two propagators that are {\it not} cut in each quadruple cut, {\it e.g.} $[x,y]$ denotes the residue with $x_k, x_{k{+}1}, x_l, x_{l{+}1}$ cut; by Global Residue Theorem (GRT) only $5$ of them are independent, which we choose to be $[x,y], [x, x_k], \cdots, [x, x_{l{+}1}]$. }, and we have
\begin{align*}
&[x,y]=\mathrm{d}\log \frac{\langle i X k l\rangle}{\langle i X j Y \rangle} \mathrm{d}\log \frac{\langle \bar{i} (j Y) \cap (i k l)\rangle }{\langle j Y k l\rangle}\,, \\
&[x, x_k]=\mathrm{d}\log \frac{\langle j Y i l\rangle}{\langle j Y k l\rangle} \mathrm{d}\log \frac{\langle i X j Y\rangle}{\langle l (i X) (j Y) (k k{+}1)\rangle}\,, \\
&I_{x, x_k}:=\tilde F(x, y, x_{k{+}1}, x_l)-\tilde F(x,y, x_{k{+}1}, x_{l{+}1}) \nonumber \\
&\quad-{\rm L}_2({l{+}1},x,y,l)+{\rm L}_2( l{+}1,x, k{+}1, l)  \\
&\quad -{\rm L}_2 ( l{+}1,y, k{+}1, l) +\log u_{l{+}1, x, y, l} \log u_{x, y, k{+}1, l{+}1}\:, \nonumber \\
&I_{x,y}:={\rm L}_2(x, k, k{+}1, l)-{\rm L}_2(x,k, k{+}1, l{+}1) \nonumber \\
&\quad -{\rm L}_2(l{+}1,x, k, l)+{\rm L}_2(l{+}1,x, k{+}1, l) \\
&\quad -{\rm L}_2( l{+}1,k, k{+}1, l)+ \log u_{l{+}1, x, k, l}\log u_{x, k, k{+}1, l{+}1}\nonumber 
\end{align*}
with ${\rm L_2}(a,b,c,d):={\rm Li_2}(1-u_{a,b,c,d})$. 
We remark that \eqref{dp_2dint} has a number of desirable properties. It is manifestly DCI and expected to evaluate to weight-$4$ functions, and one can check that it remains finite even for special cases such as $j=k{+}1$ or $i=l{+}1$. Moreover, for the generic case we have $4$ four-mass boxes involved, which depend on square roots $\Delta(x, y, k, l):=\sqrt{(1-u-v)^{2}-4uv}$ ($u,v$ are defined above for these four points) {\it etc.}, and after integrating over $x,y$, each needs to be evaluated at endpoints $x=x_i, x_{i{+}1}$ (similarly for $y$); thus the result must contain the $16$ square roots  $\Delta(i, j, k, l)$, $\Delta(i{+}1, j, k, l)$, $\Delta(i, j{+}1, k, l)$, $\Delta(i{+}1, j{+}1, k,l)$ {\it etc.} as expected.

\section{Rationalization: uniform transcendentality, algebraic words and their cancellation}\label{rationalization}

Had there been no square root in $I_{\rm hex}$, it would have been straightforward to perform the two-fold integrations in \eqref{dpWL}.  In addition to square roots in $\tilde F$'s, what is worse is the presence of $\gamma$'s which makes it even obscure that the answer must be pure! It turns out that these issues are resolved by ``rationalizing" the square roots, which have been exploited in the ${\bar Q}$ calculation~\cite{Zhang:2019vnm, He:2020vob}. The idea is very simple: we make change of variables such that there is no square root in $I_{\rm hex}$, then the integral can be performed {\it e.g.} at the symbol level using the algorithm given in~\cite{CaronHuot:2011kk}, and square roots only appear via integration domains. Let us consider any of the $4$ four-mass boxes $\tilde F(x(\tau), y(\tau'), *, *)$, where the square roots are contained in $z(\tau, \tau')$ and $\bar{z}(\tau, \tau')$; the $\mathrm{d}\log$ forms we write indicate that we should do $\tau$ integral first. We make change of variable from $\tau$ to $z(\tau)$ (suppressing the dependence on $\tau'$). As for the $\bar{z}$, note that there exist $a$ and $b$, which depend on kinematics and $\tau'$, but independent of $\tau$, such that $a u(\tau)+ b v(\tau)=1$. This allows us to relate $\bar{z}(\tau)$ to $z(\tau)$ by a M\"{o}bius transformation $\bar{z}=\Lambda(z):=\frac{b z+ (1-b)}{(a+b)z-b}$. 

Something remarkable happens at this stage: the prefactor $\gamma$, together with $\mathrm{d}\log$ forms depending on $\tau$, becomes a beautiful $\mathrm{d}\log$ of a rational function of $z(\tau)$! The $\tau$-integral of a four-mass box function becomes: 
\begin{align} 
 &\int_{z(0)}^{z(\infty)} \mathrm{d}\log \frac{z-w}{z-\Lambda(w)} \biggl({\rm Li}_2 (1-z)-{\rm Li}_2 (1-\Lambda(z))\nonumber \\
  &\quad +\frac{1}{2}\log\frac{z}{\Lambda(z)} \log\bigl((1-z)(1-\Lambda(z))\bigr)\biggr)\:,\label{rarepofw2} 
\end{align}
for some $w$ and $\Lambda(w)$, both of which are independent of $\tau$. At this stage, it becomes obvious that $I_{\rm dp}$ is represented as $2$-fold $\mathrm{d}\log$ integrals of weight-$2$ pure functions. 

In this form, one can perform the $\tau$-integration directly, and it suffices to give the part of the symbol only involving square roots, which depends on $\tau'$. The algebraic part of the above integral \eqref{rarepofw2} gives a beautiful weight-3 ``algebraic word" (of the $\tau'$-integrand):
$$
\frac 14 \left.\left(u \otimes \frac{1-\bar z}{1-z} + v\otimes \frac{z}{\bar z}\right) \otimes \frac{(z-w)(\Lambda(z)-\Lambda(w))}{(\Lambda(z)-w)(z-\Lambda(w))}\right\vert^{\tau=\infty}_{\tau=0}\,,
$$
where we evaluate the symbol at $\tau=\infty$, minus that at $\tau=0$, which results in square roots $\Delta(x_i, y, *, *)$ and $\Delta(x_{i{+}1}, y, *, *)$, and note that the first two-entries are exactly the symbol of four-mass box, $F(u,v)$. As one can easily check that these weight-$3$ algebraic words cancel in the difference $I_{\rm dp} (i, j, k, l)-I_{\rm dp} (j, k, l, i)$: all square roots drop out already at the $\tau'$-integrand level!


Next we perform the $\tau'$-integration, and we need to rationalize the square roots in $\tau'$ of the above algebraic words. We emphasize a major difference between this step and the previous one from weight-$2$ to $3$: the $d\log$ factors are manifestly rational due to the absence of $\gamma$ factor, thus after we change variable from $\tau'$ to $z(\tau')$, the arguments of $\mathrm{d}\log$'s are given by {\it products} of the form $(z-w)(z-\Lambda'(w))$ rather than ratios. This has an immediate consequence that the last entries of the resulting symbol are free of any four-mass square roots~\footnote{We still need to perform ``rationalization'' and encounter spurious square roots of kinematics from $w$ , but they all nicely cancel in the final answer as expected.}. In the end, we obtain a remarkably compact expression for ``algebraic words'' of the final answer: the first two entries are given by (the symbol of) four-mass boxes, the third entry given by algebraic letters, and the last entry rational ones. Since there are $16$ square roots, $\Delta(a,b,c,d)$ for $a:=i+\sigma_1, b:=j+\sigma_2, c:=k+\sigma_3, d:=l+\sigma_4$ with $\sigma=0,1$, the algebraic part of the symbol of $I_{\rm dp}(i,j,k,l)$ can be written as an alternating sum of $16$ terms:
{\small \begin{equation}
\hspace{-2ex}\sum_{\sigma_a \in \{0, 1\}} (-)^{\sum \sigma}
S[F({i{+}\sigma_1}, 
{j{+}\sigma_2}, {k{+}\sigma_3}, {l{+}\sigma_4})] \otimes W^{i,j,k,l}_{\sigma_1, \cdots, \sigma_4}
\end{equation}}where each term is characterized by a four-mass box $F(a,b,c,d)$; 
it is accompanied by last two-entries denoted as $W^{i,j,k,l}_{\sigma_1, \cdots, \sigma_4}$, which contains the same square root $\Delta(a,b,c,d)$ and depends on $x_a, \cdots, x_d$ and $i,j,k,l$.
{\small \begin{align}
&W^{i,j,k,l}_{a-i, \cdots, d-l}
=\chi_{a,b,c,d}^{j,k} \otimes  \frac{\langle x_a \,jk \rangle \langle x_b \,i l \rangle}{\langle x_a \,jl \rangle \langle x_b\, i k \rangle} + {\rm cyclic}
\label{algwords}\\
&+\frac 1 2 \left(
\frac{\bar{z}(1{-}z)}{z(1{-}\bar{z})}
\prod \chi \right)\otimes \frac{\langle x_a \,j l\rangle \langle x_b\,i k\rangle\langle x_c \,j l \rangle\langle x_d 
\,i k\rangle}{\langle x_a\, kl\rangle \langle x_b \,i l\rangle\langle x_c \,ij \rangle\langle x_d \,jk\rangle}\nonumber
\end{align}}where the first four terms are given by cyclic rotation in $i,j,k,l$ (thus also in $a,b,c,d$), and in the last term,  both $\frac{\bar{z}(1{-}z)}{z(1{-}\bar{z})}$ and the product $\prod \chi$ are cyclic invariant; 
the $4$ new algebraic letters are given by
$$
\chi_{a,b,c,d}^{j,k}:=\left(\frac{\frac{\langle x_a x_b \rangle \langle x_d\,jk \rangle}{\langle x_d x_b \rangle \langle x_a \,jk \rangle}-z_{a,b,c,d}}{\frac{\langle x_a x_b\rangle \langle x_d \,jk \rangle}{\langle x_d x_b \rangle \langle x_a \,jk \rangle}-\bar{z}_{a,b,c,d}}\right)
$$
and cyclic images $\chi_{b,c,d,a}^{k,l}$, $\chi_{c,d,a,b}^{l,i}$ and $\chi_{d,a,b,c}^{i,j}$. Note that the algebraic letters are special multiplicative combinations of those found for two-loop NMHV in~\cite{He:2020vob}.

The way we present $W$ makes manifest a nice symmetry , $W^{i,j,k,l}_{\sigma_1, \sigma_2, \sigma_3, \sigma_4}=W^{j,k,l,i}_{\sigma_2, \sigma_3, \sigma_4, \sigma_1}$, which guarantees that all square roots cancel in $I_{\rm dp}(i,j,k,l)-I_{\rm dp}(j,k,l,i)$, as we have seen at the level of weight-3 integrands. It is even more interesting to see how square roots also drop out for two-loop MHV amplitudes (given by a cyclic sum of all $I_{\rm dp}$). To see this, we collect algebraic words for a given square root: it is easy to see that $16$ $I_{\rm dp}$ contribute, 
and the result is given by the tensor product of $S[F(x_a, \cdots, x_d)]$ and the combination
\begin{equation}
\sum_{\sigma_a \in \{0, 1\}} W^{a-\sigma_1, \cdots, d-\sigma_4}_{\sigma_1, \cdots, \sigma_4} (x_a, x_b, x_c, x_d) \:.
\end{equation} 
Quite nicely, we find it vanishes, which guarantees the absence of square roots for two-loop MHV amplitudes.

\section{Final Results and Checks} \label{dataandcheck}
In addition to the algebraic part, we also compute the remaining part that is free of any square roots; the computation of the symbol can be done trivially, as long as we apply the integration-rule consistently to the complete weight-$3$ symbol including the ``algebraic words" and the rest~\footnote{Note that when converting ${\rm d}\log$'s of $\tau'$ into those of $z(\tau')$, we omitted ``constants'' that are independent of $\tau'$; it is crucial to also drop the same constants for ${\rm d}\log$'s in the rational part, since one can only ignore such constants for the entire, integrable symbol.}. We record the symbol for $I_{\rm dp} (1,4,7,10)$ with $n=12$ in the ancillary file {\bf result.nb}. 
Remarkably, we find that the complete symbol can be written in a compact form by organizing it using $8+16$ combinations of last entries with manifest symmetries. Equivalently, we express its total differential ${\rm d} I_{\rm dp}(i,j,k,l)$ as
{\small \begin{align}\label{complete}
&\frac 1 2 R^{\bar i}_{j{-}1 j} {\rm d}\log \frac{\langle i (i{-}1 i{+}1) (j{-}1 j) (k l) \rangle}{\langle \bar{i} j\rangle \langle j{-}1 j k l \rangle}+ M^{i k l}_{j{-}1 j} {\rm d}\log \frac{\langle i j{-}1 j k\rangle}{\langle j{-}1 j k l\rangle}
\nonumber
\\
&- (j{-}1 j \leftrightarrow j j{+}1) + (\bar{i} \leftrightarrow \bar{j})+  (\bar{k} \leftrightarrow \bar{l})+ (i\,j \leftrightarrow k\,l)
\end{align}}where only {\it two} independent weight-$3$ DCI functions, $R^{\bar i}_{j{-}1 j}$ and $M^{i k l}_{j{-}1 j}$ are needed; each relabelling applies to all previous terms, giving $8$ and $16$ images of these functions, respectively. The algebraic words \eqref{algwords} contribute to the symbol of $M$ only, and that of $R$ is rational; we present both symbols in the appendix. 

Let us briefly summarize rational letters appearing in different entries of the symbol (also listed in {\bf result.nb}). As seen from \eqref{complete}, there are $36$ letters for the {\bf last entry}: $24$ Pl\"{u}cker coordinates accompanying the symbol of $F$, which are $\langle a{-}1 a j k \rangle$, $\langle a{-}1 a k l\rangle$, $\langle a{-}1 a j l\rangle $ plus cyclic in $i,j,k,l$, and $12$ accompanying that of $R$: $8$ of the form $\langle i (i{-}1 i{+}1) (b{-}1 b) (k l )\rangle $ with $b=j$ or $j{+}1$, plus cyclic in $i,j,k,l$, and $\langle i \bar{j} \rangle, \langle j \bar{i}\rangle, \langle k \bar{l}\rangle, \langle l \bar{k}\rangle$ (they remain as last entries of MHV amplitudes). Next we describe rational letters in the first three entries in order. 

Only physical discontinuities can appear for the {\bf first entry} (note in our notation $a=i,i{+}1$, $b=j, j{+}1$, $c=k, k{+}1$, $d=l, l{+}1$, thus there are ${4 \choose 2}\times 4=24$ in total):
$\langle a{-}1 a\,b{-}1 b\rangle$, $\langle a{-}1 a\,c{-}1 c\rangle$, $\langle a{-}1 a\,d{-}1 d\rangle$, $\langle b{-}1 b\,c{-}1 c\rangle$, $\langle b{-}1 b\,d{-}1 d\rangle$, $\langle c{-}1 c\,d{-}1 d\rangle$; 
in the rational alphabet, we additionally have the following $60$ letters for the {\bf second entry}:
${4 \choose 2}\times 2=12$ of the form $\langle i \bar{j}\rangle$ {\it etc.} and $4\times 4\times {3 \choose 2}=48$ of the form $\langle i (i{-}1\, i{+}1) (b{-}1 \,b) (c{-}1\, c) \rangle$, $\langle i (i{-}1 i{+}1) (b{-}1, b) (d{-}1 d)\rangle$, $\langle i (i{-}1 i{+}1) (c{-}1 c) (d{-}1) d\rangle$ (plus cyclic in $i,j,k,l$). These first two entries agree with the prediction in~\cite{Dennen:2015bet} from Landau analysis, and they are consistent with the general prediction that first two entries of rational part can only be ${\rm Li}_2(1-u)$ or $\log u \log v$ with $u,v$ made from physical discontinuities; these dilogarithm functions can be viewed as degeneration of four-mass boxes $F(u,v)$ (whose symbols form the first two entries of algebraic words). 

We find 140 rational letters appearing in the {\bf third entry}, and it suffices to list the only two types that have not appeared in other entries: $4\times 8=32$ letters of the form $\langle i (b{-}1\, b) (c{-}1\, c) (d{-}1 \,d)\rangle$, 
and $4\times 4=16$ of the form $\langle (\bar{i}) \cap (i b{-}1 b) \cap (\bar{k}) \cap (k d{-}1 d) \rangle$ 
{\it etc.}. In total, we find that the alphabet consists of $164$ rational letters: 
$60$ Pl\"{u}cker and $104$ non-Pl\"{u}cker rational letters as described above.

We have performed thorough checks on our result, where an important check mentioned above is the physical first-entry condition (as well as that for first two entries). One can easily check that the symbol is DCI, and as shown in~\eqref{complete} it 
is symmetric in exchanging $(i,j)$ with $(l,k)$ and in simultaneous exchange $i\leftrightarrow j$ and $k\leftrightarrow l$, as well as anti-symmetric under $i{-}1\leftrightarrow i{+}1$ {\it etc.}. 
A non-trivial check is to see that the complete symbol is integrable. Moreover, given the most generic $I_{\rm dp} (i,j,k,l)$, it is important that any finite degeneration of the integral remains well-defined. This happens when $j=i{+}2$ (similarly $l=k{+}2$), or $k=j{+}2$ 
(similarly between $l$ and $i$), 
and they can be viewed as (multiple) collinear limits of the original integral. We have checked that in all these cases the symbol remains finite, which also gives results for these special cases. For example, $I_{\rm dp} (1,3,5,7)$ for $n=8$ can be obtained from the generic case by taking $4$ collinear limits; the symbol, recorded in {\bf result.nb} as well, contains $2$ square roots and $108$ rational letters. Last but not least, we take the difference $I_{\rm dp}(1,3,5,7)-I_{\rm dp}(3,5,7,1)$ and find perfect agreement with the component $\chi_1 \chi_3 \chi_5 \chi_7$ of $n=8$ NMHV amplitudes from ${\bar Q}$ calculation~\cite{Zhang:2019vnm}.  
\section{Conclusions and Outlook}
We have computed the symbol of all finite double-pentagon integrals $I_{\rm dp} (i,j,k,l)$ (with $j>i{+}1$ and $l>k{+}1$), which also amounts to all-multiplicity, Feynman-integral computation of the finite part of two-loop MHV amplitudes, and all non-adjacent $\chi_i \chi_j \chi_k \chi_l$ components of two-loop NMHV amplitudes. The alphabet consists of $96$ algebraic letters ($6$ for each of the $16$ square roots), and $164$ rational letters, and we see not only desirable physical conditions on first two entries, but also more interesting patterns for the complete symbol. 
The compact expression \eqref{complete} with the symbol of $R$ and $M$ in the appendix deserves further investigations, which also gives a compact formula for the square-root-free symbol of two-loop NMHV components. It would be interesting to determine the weight-$3$ functions $R$ and $M$, which may have interesting physical meaning themselves. Of course it would also be nice to upgrade the symbol to weight-$4$ functions (one possibility being the bootstrap method along the line of~\cite{Henn:2018cdp}). Another important issue is to compute divergent integrals using certain regularization, which would allow us to compare with the remainder function computed in~\cite{CaronHuot:2011ky, CaronHuot:2011kk}. 

More generally, the method using WL picture has been applied to a wide range of multi-loop integrals~\cite{He:2020uxy} but for generic cases similar rationalizations are again needed. It is straightforward to compute other important two-loop integrals with a chiral-pentagon loop, such as the other type of double-pentagon integrals needed for NMHV amplitudes~\cite{ArkaniHamed:2010kv} and penta-box integrals {\it etc.}~\cite{Bourjaily:2015jna}, which would give complete two-loop NMHV amplitudes and components of N${}^2$MHV ones. Three-loop integrals such as those for MHV amplitudes are also within reach, though they certainly require more efforts. It would be interesting to see how the method can be connected to differential equations in~\cite{Drummond:2010cz,Henn:2013pwa} and ${\bar Q}$ method~\cite{CaronHuot:2011kk} for Feynman integrals and scattering amplitudes. Our method for rationalizations is universal and applies to ${\bar Q}$ calculation for amplitudes: the step from weight-$2$ to weight-$3$ with non-trivial prefactor $\gamma$, resembles the computation of two-loop NMHV amplitudes from one-loop N${}^2$MHV ones~\cite{Zhang:2019vnm, He:2020uxy} (so do resulting algebraic words); the simpler step from weight-$3$ to weight-$4$ can be used for computing three-loop MHV amplitudes from two-loop NMHV ones, and in particular predicting the structure of algebraic words in the symbol, which will be present in a future work. 

Perhaps even more interesting are structural questions regarding our method and result. For example, it would be highly desirable to deduce general information about the symbol and functions of integrals in WL ${\rm d}\log$ representation without explicit computations as we did for certain ladders in~\cite{He:2020uxy}. It would be extremely interesting to study possible geometric structures of these WL integrals, with hints from various ${\rm d}\log$ forms we found; for $I_{\rm dp}(i,j,k,l)$, a more symmetric computation expresses it as four-fold integrals of algebraic functions (up to possible boundary terms with less integrations of polylogarithms), which may reveal some underlying geometries for these integrals and beyond. 

\begin{acknowledgments}
We would like to thank Nima Arkani-Hamed, Yichao Tang and Yang Zhang for inspiring discussions. SH's research is supported in part by the Key Research Program of Frontier Sciences of CAS under Grant No. QYZDBSSW-SYS014, Peng Huanwu center under Grant No. 12047503 and National Natural Science Foundation of China under Grant No. 12047503. CZ was supported in part by the ERC starting grant 757978 and grant 00025445 from the Villum Fonden.
\end{acknowledgments}



\bibliography{dp_WL}

\providecommand{\noopsort}[1]{}\providecommand{\singleletter}[1]{#1}%
\begin{thebibliography}{59}%
\makeatletter
\providecommand \@ifxundefined [1]{%
 \@ifx{#1\undefined}
}%
\providecommand \@ifnum [1]{%
 \ifnum #1\expandafter \@firstoftwo
 \else \expandafter \@secondoftwo
 \fi
}%
\providecommand \@ifx [1]{%
 \ifx #1\expandafter \@firstoftwo
 \else \expandafter \@secondoftwo
 \fi
}%
\providecommand \natexlab [1]{#1}%
\providecommand \enquote  [1]{``#1''}%
\providecommand \bibnamefont  [1]{#1}%
\providecommand \bibfnamefont [1]{#1}%
\providecommand \citenamefont [1]{#1}%
\providecommand \href@noop [0]{\@secondoftwo}%
\providecommand \href [0]{\begingroup \@sanitize@url \@href}%
\providecommand \@href[1]{\@@startlink{#1}\@@href}%
\providecommand \@@href[1]{\endgroup#1\@@endlink}%
\providecommand \@sanitize@url [0]{\catcode `\\12\catcode `\$12\catcode
  `\&12\catcode `\#12\catcode `\^12\catcode `\_12\catcode `\%12\relax}%
\providecommand \@@startlink[1]{}%
\providecommand \@@endlink[0]{}%
\providecommand \url  [0]{\begingroup\@sanitize@url \@url }%
\providecommand \@url [1]{\endgroup\@href {#1}{\urlprefix }}%
\providecommand \urlprefix  [0]{URL }%
\providecommand \Eprint [0]{\href }%
\providecommand \doibase [0]{http://dx.doi.org/}%
\providecommand \selectlanguage [0]{\@gobble}%
\providecommand \bibinfo  [0]{\@secondoftwo}%
\providecommand \bibfield  [0]{\@secondoftwo}%
\providecommand \translation [1]{[#1]}%
\providecommand \BibitemOpen [0]{}%
\providecommand \bibitemStop [0]{}%
\providecommand \bibitemNoStop [0]{.\EOS\space}%
\providecommand \EOS [0]{\spacefactor3000\relax}%
\providecommand \BibitemShut  [1]{\csname bibitem#1\endcsname}%
\let\auto@bib@innerbib\@empty
\bibitem [{\citenamefont {Arkani-Hamed}\ \emph {et~al.}(2011)\citenamefont
  {Arkani-Hamed}, \citenamefont {Bourjaily}, \citenamefont {Cachazo},
  \citenamefont {Caron-Huot},\ and\ \citenamefont
  {Trnka}}]{ArkaniHamed:2010kv}%
  \BibitemOpen
  \bibfield  {author} {\bibinfo {author} {\bibfnamefont {N.}~\bibnamefont
  {Arkani-Hamed}}, \bibinfo {author} {\bibfnamefont {J.~L.}\ \bibnamefont
  {Bourjaily}}, \bibinfo {author} {\bibfnamefont {F.}~\bibnamefont {Cachazo}},
  \bibinfo {author} {\bibfnamefont {S.}~\bibnamefont {Caron-Huot}}, \ and\
  \bibinfo {author} {\bibfnamefont {J.}~\bibnamefont {Trnka}},\ }\href
  {\doibase 10.1007/JHEP01(2011)041} {\bibfield  {journal} {\bibinfo  {journal}
  {JHEP}\ }\textbf {\bibinfo {volume} {01}},\ \bibinfo {pages} {041} (\bibinfo
  {year} {2011})},\ \Eprint {http://arxiv.org/abs/1008.2958} {arXiv:1008.2958
  [hep-th]} \BibitemShut {NoStop}%
\bibitem [{\citenamefont {Arkani-Hamed}\ \emph {et~al.}(2016)\citenamefont
  {Arkani-Hamed}, \citenamefont {Bourjaily}, \citenamefont {Cachazo},
  \citenamefont {Goncharov}, \citenamefont {Postnikov},\ and\ \citenamefont
  {Trnka}}]{ArkaniHamed:2012nw}%
  \BibitemOpen
  \bibfield  {author} {\bibinfo {author} {\bibfnamefont {N.}~\bibnamefont
  {Arkani-Hamed}}, \bibinfo {author} {\bibfnamefont {J.~L.}\ \bibnamefont
  {Bourjaily}}, \bibinfo {author} {\bibfnamefont {F.}~\bibnamefont {Cachazo}},
  \bibinfo {author} {\bibfnamefont {A.~B.}\ \bibnamefont {Goncharov}}, \bibinfo
  {author} {\bibfnamefont {A.}~\bibnamefont {Postnikov}}, \ and\ \bibinfo
  {author} {\bibfnamefont {J.}~\bibnamefont {Trnka}},\ }\href
  {https://inspirehep.net/record/1208741/files/arXiv:1212.5605.pdf} {\emph
  {\bibinfo {title} {{Grassmannian Geometry of Scattering Amplitudes}}}}\
  (\bibinfo  {publisher} {Cambridge University Press},\ \bibinfo {year}
  {2016})\ \Eprint {http://arxiv.org/abs/1212.5605} {arXiv:1212.5605 [hep-th]}
  \BibitemShut {NoStop}%
\bibitem [{\citenamefont {Arkani-Hamed}\ and\ \citenamefont
  {Trnka}(2014)}]{Arkani-Hamed:2013jha}%
  \BibitemOpen
  \bibfield  {author} {\bibinfo {author} {\bibfnamefont {N.}~\bibnamefont
  {Arkani-Hamed}}\ and\ \bibinfo {author} {\bibfnamefont {J.}~\bibnamefont
  {Trnka}},\ }\href {\doibase 10.1007/JHEP10(2014)030} {\bibfield  {journal}
  {\bibinfo  {journal} {JHEP}\ }\textbf {\bibinfo {volume} {10}},\ \bibinfo
  {pages} {030} (\bibinfo {year} {2014})},\ \Eprint
  {http://arxiv.org/abs/1312.2007} {arXiv:1312.2007 [hep-th]} \BibitemShut
  {NoStop}%
\bibitem [{\citenamefont {Dixon}\ \emph {et~al.}(2011)\citenamefont {Dixon},
  \citenamefont {Drummond},\ and\ \citenamefont {Henn}}]{Dixon:2011pw}%
  \BibitemOpen
  \bibfield  {author} {\bibinfo {author} {\bibfnamefont {L.~J.}\ \bibnamefont
  {Dixon}}, \bibinfo {author} {\bibfnamefont {J.~M.}\ \bibnamefont {Drummond}},
  \ and\ \bibinfo {author} {\bibfnamefont {J.~M.}\ \bibnamefont {Henn}},\
  }\href {\doibase 10.1007/JHEP11(2011)023} {\bibfield  {journal} {\bibinfo
  {journal} {JHEP}\ ,\ \bibinfo {pages} {023}} (\bibinfo {year} {2011})},\
  \Eprint {http://arxiv.org/abs/1108.4461} {arXiv:1108.4461 [hep-th]}
  \BibitemShut {NoStop}%
\bibitem [{\citenamefont {Dixon}\ \emph {et~al.}(2014)\citenamefont {Dixon},
  \citenamefont {Drummond}, \citenamefont {Duhr}, \citenamefont {von Hippel},\
  and\ \citenamefont {Pennington}}]{Dixon:2014xca}%
  \BibitemOpen
  \bibfield  {author} {\bibinfo {author} {\bibfnamefont {L.~J.}\ \bibnamefont
  {Dixon}}, \bibinfo {author} {\bibfnamefont {J.~M.}\ \bibnamefont {Drummond}},
  \bibinfo {author} {\bibfnamefont {C.}~\bibnamefont {Duhr}}, \bibinfo {author}
  {\bibfnamefont {M.}~\bibnamefont {von Hippel}}, \ and\ \bibinfo {author}
  {\bibfnamefont {J.}~\bibnamefont {Pennington}},\ }\href {\doibase
  10.22323/1.211.0077} {\bibfield  {journal} {\bibinfo  {journal} {PoS}\
  }\textbf {\bibinfo {volume} {LL2014}},\ \bibinfo {pages} {077} (\bibinfo
  {year} {2014})},\ \Eprint {http://arxiv.org/abs/1407.4724} {arXiv:1407.4724
  [hep-th]} \BibitemShut {NoStop}%
\bibitem [{\citenamefont {Dixon}\ and\ \citenamefont {von
  Hippel}(2014)}]{Dixon:2014iba}%
  \BibitemOpen
  \bibfield  {author} {\bibinfo {author} {\bibfnamefont {L.~J.}\ \bibnamefont
  {Dixon}}\ and\ \bibinfo {author} {\bibfnamefont {M.}~\bibnamefont {von
  Hippel}},\ }\href {\doibase 10.1007/JHEP10(2014)065} {\bibfield  {journal}
  {\bibinfo  {journal} {JHEP}\ }\textbf {\bibinfo {volume} {10}},\ \bibinfo
  {pages} {065} (\bibinfo {year} {2014})},\ \Eprint
  {http://arxiv.org/abs/1408.1505} {arXiv:1408.1505 [hep-th]} \BibitemShut
  {NoStop}%
\bibitem [{\citenamefont {Drummond}\ \emph {et~al.}(2015)\citenamefont
  {Drummond}, \citenamefont {Papathanasiou},\ and\ \citenamefont
  {Spradlin}}]{Drummond:2014ffa}%
  \BibitemOpen
  \bibfield  {author} {\bibinfo {author} {\bibfnamefont {J.~M.}\ \bibnamefont
  {Drummond}}, \bibinfo {author} {\bibfnamefont {G.}~\bibnamefont
  {Papathanasiou}}, \ and\ \bibinfo {author} {\bibfnamefont {M.}~\bibnamefont
  {Spradlin}},\ }\href {\doibase 10.1007/JHEP03(2015)072} {\bibfield  {journal}
  {\bibinfo  {journal} {JHEP}\ }\textbf {\bibinfo {volume} {03}},\ \bibinfo
  {pages} {072} (\bibinfo {year} {2015})},\ \Eprint
  {http://arxiv.org/abs/1412.3763} {arXiv:1412.3763 [hep-th]} \BibitemShut
  {NoStop}%
\bibitem [{\citenamefont {Dixon}\ \emph {et~al.}(2016)\citenamefont {Dixon},
  \citenamefont {von Hippel},\ and\ \citenamefont {McLeod}}]{Dixon:2015iva}%
  \BibitemOpen
  \bibfield  {author} {\bibinfo {author} {\bibfnamefont {L.~J.}\ \bibnamefont
  {Dixon}}, \bibinfo {author} {\bibfnamefont {M.}~\bibnamefont {von Hippel}}, \
  and\ \bibinfo {author} {\bibfnamefont {A.~J.}\ \bibnamefont {McLeod}},\
  }\href {\doibase 10.1007/JHEP01(2016)053} {\bibfield  {journal} {\bibinfo
  {journal} {JHEP}\ }\textbf {\bibinfo {volume} {01}},\ \bibinfo {pages} {053}
  (\bibinfo {year} {2016})},\ \Eprint {http://arxiv.org/abs/1509.08127}
  {arXiv:1509.08127 [hep-th]} \BibitemShut {NoStop}%
\bibitem [{\citenamefont {Caron-Huot}\ \emph {et~al.}(2016)\citenamefont
  {Caron-Huot}, \citenamefont {Dixon}, \citenamefont {McLeod},\ and\
  \citenamefont {von Hippel}}]{Caron-Huot:2016owq}%
  \BibitemOpen
  \bibfield  {author} {\bibinfo {author} {\bibfnamefont {S.}~\bibnamefont
  {Caron-Huot}}, \bibinfo {author} {\bibfnamefont {L.~J.}\ \bibnamefont
  {Dixon}}, \bibinfo {author} {\bibfnamefont {A.}~\bibnamefont {McLeod}}, \
  and\ \bibinfo {author} {\bibfnamefont {M.}~\bibnamefont {von Hippel}},\
  }\href {\doibase 10.1103/PhysRevLett.117.241601} {\bibfield  {journal}
  {\bibinfo  {journal} {Phys. Rev. Lett.}\ }\textbf {\bibinfo {volume} {117}},\
  \bibinfo {pages} {241601} (\bibinfo {year} {2016})},\ \Eprint
  {http://arxiv.org/abs/1609.00669} {arXiv:1609.00669 [hep-th]} \BibitemShut
  {NoStop}%
\bibitem [{\citenamefont {Caron-Huot}(2011{\natexlab{a}})}]{CaronHuot:2011ky}%
  \BibitemOpen
  \bibfield  {author} {\bibinfo {author} {\bibfnamefont {S.}~\bibnamefont
  {Caron-Huot}},\ }\href {\doibase 10.1007/JHEP12(2011)066} {\bibfield
  {journal} {\bibinfo  {journal} {JHEP}\ }\textbf {\bibinfo {volume} {12}},\
  \bibinfo {pages} {066} (\bibinfo {year} {2011}{\natexlab{a}})},\ \Eprint
  {http://arxiv.org/abs/1105.5606} {arXiv:1105.5606 [hep-th]} \BibitemShut
  {NoStop}%
\bibitem [{\citenamefont {He}\ \emph {et~al.}(2020{\natexlab{a}})\citenamefont
  {He}, \citenamefont {Li},\ and\ \citenamefont {Zhang}}]{Zhang:2019vnm}%
  \BibitemOpen
  \bibfield  {author} {\bibinfo {author} {\bibfnamefont {S.}~\bibnamefont
  {He}}, \bibinfo {author} {\bibfnamefont {Z.}~\bibnamefont {Li}}, \ and\
  \bibinfo {author} {\bibfnamefont {C.}~\bibnamefont {Zhang}},\ }\href
  {\doibase 10.1103/PhysRevD.101.061701} {\bibfield  {journal} {\bibinfo
  {journal} {Phys. Rev. D}\ }\textbf {\bibinfo {volume} {101}},\ \bibinfo
  {pages} {061701} (\bibinfo {year} {2020}{\natexlab{a}})},\ \Eprint
  {http://arxiv.org/abs/1911.01290} {arXiv:1911.01290 [hep-th]} \BibitemShut
  {NoStop}%
\bibitem [{\citenamefont {He}\ \emph {et~al.}(2020{\natexlab{b}})\citenamefont
  {He}, \citenamefont {Li},\ and\ \citenamefont {Zhang}}]{He:2020vob}%
  \BibitemOpen
  \bibfield  {author} {\bibinfo {author} {\bibfnamefont {S.}~\bibnamefont
  {He}}, \bibinfo {author} {\bibfnamefont {Z.}~\bibnamefont {Li}}, \ and\
  \bibinfo {author} {\bibfnamefont {C.}~\bibnamefont {Zhang}},\ }\href@noop {}
  {\  (\bibinfo {year} {2020}{\natexlab{b}})},\ \Eprint
  {http://arxiv.org/abs/2009.11471} {arXiv:2009.11471 [hep-th]} \BibitemShut
  {NoStop}%
\bibitem [{\citenamefont {Bourjaily}\ \emph {et~al.}(2018)\citenamefont
  {Bourjaily}, \citenamefont {McLeod}, \citenamefont {von Hippel},\ and\
  \citenamefont {Wilhelm}}]{Bourjaily:2018aeq}%
  \BibitemOpen
  \bibfield  {author} {\bibinfo {author} {\bibfnamefont {J.~L.}\ \bibnamefont
  {Bourjaily}}, \bibinfo {author} {\bibfnamefont {A.~J.}\ \bibnamefont
  {McLeod}}, \bibinfo {author} {\bibfnamefont {M.}~\bibnamefont {von Hippel}},
  \ and\ \bibinfo {author} {\bibfnamefont {M.}~\bibnamefont {Wilhelm}},\ }\href
  {\doibase 10.1007/JHEP08(2018)184} {\bibfield  {journal} {\bibinfo  {journal}
  {JHEP}\ }\textbf {\bibinfo {volume} {08}},\ \bibinfo {pages} {184} (\bibinfo
  {year} {2018})},\ \Eprint {http://arxiv.org/abs/1805.10281} {arXiv:1805.10281
  [hep-th]} \BibitemShut {NoStop}%
\bibitem [{\citenamefont {Henn}\ \emph {et~al.}(2018)\citenamefont {Henn},
  \citenamefont {Herrmann},\ and\ \citenamefont
  {Parra-Martinez}}]{Henn:2018cdp}%
  \BibitemOpen
  \bibfield  {author} {\bibinfo {author} {\bibfnamefont {J.}~\bibnamefont
  {Henn}}, \bibinfo {author} {\bibfnamefont {E.}~\bibnamefont {Herrmann}}, \
  and\ \bibinfo {author} {\bibfnamefont {J.}~\bibnamefont {Parra-Martinez}},\
  }\href {\doibase 10.1007/JHEP10(2018)059} {\bibfield  {journal} {\bibinfo
  {journal} {JHEP}\ }\textbf {\bibinfo {volume} {10}},\ \bibinfo {pages} {059}
  (\bibinfo {year} {2018})},\ \Eprint {http://arxiv.org/abs/1806.06072}
  {arXiv:1806.06072 [hep-th]} \BibitemShut {NoStop}%
\bibitem [{\citenamefont {Herrmann}\ and\ \citenamefont
  {Parra-Martinez}(2019)}]{Herrmann:2019upk}%
  \BibitemOpen
  \bibfield  {author} {\bibinfo {author} {\bibfnamefont {E.}~\bibnamefont
  {Herrmann}}\ and\ \bibinfo {author} {\bibfnamefont {J.}~\bibnamefont
  {Parra-Martinez}},\ }\href@noop {} {\  (\bibinfo {year} {2019})},\ \Eprint
  {http://arxiv.org/abs/1909.04777} {arXiv:1909.04777 [hep-th]} \BibitemShut
  {NoStop}%
\bibitem [{\citenamefont {Alday}\ and\ \citenamefont
  {Maldacena}(2007{\natexlab{a}})}]{Alday:2007hr}%
  \BibitemOpen
  \bibfield  {author} {\bibinfo {author} {\bibfnamefont {L.~F.}\ \bibnamefont
  {Alday}}\ and\ \bibinfo {author} {\bibfnamefont {J.~M.}\ \bibnamefont
  {Maldacena}},\ }\href {\doibase 10.1088/1126-6708/2007/06/064} {\bibfield
  {journal} {\bibinfo  {journal} {JHEP}\ }\textbf {\bibinfo {volume} {06}},\
  \bibinfo {pages} {064} (\bibinfo {year} {2007}{\natexlab{a}})},\ \Eprint
  {http://arxiv.org/abs/0705.0303} {arXiv:0705.0303 [hep-th]} \BibitemShut
  {NoStop}%
\bibitem [{\citenamefont {Alday}\ and\ \citenamefont
  {Maldacena}(2007{\natexlab{b}})}]{Alday:2007he}%
  \BibitemOpen
  \bibfield  {author} {\bibinfo {author} {\bibfnamefont {L.~F.}\ \bibnamefont
  {Alday}}\ and\ \bibinfo {author} {\bibfnamefont {J.}~\bibnamefont
  {Maldacena}},\ }\href {\doibase 10.1088/1126-6708/2007/11/068} {\bibfield
  {journal} {\bibinfo  {journal} {JHEP}\ }\textbf {\bibinfo {volume} {11}},\
  \bibinfo {pages} {068} (\bibinfo {year} {2007}{\natexlab{b}})},\ \Eprint
  {http://arxiv.org/abs/0710.1060} {arXiv:0710.1060 [hep-th]} \BibitemShut
  {NoStop}%
\bibitem [{\citenamefont {Alday}\ and\ \citenamefont
  {Maldacena}(2009)}]{Alday:2009yn}%
  \BibitemOpen
  \bibfield  {author} {\bibinfo {author} {\bibfnamefont {L.~F.}\ \bibnamefont
  {Alday}}\ and\ \bibinfo {author} {\bibfnamefont {J.}~\bibnamefont
  {Maldacena}},\ }\href {\doibase 10.1088/1126-6708/2009/11/082} {\bibfield
  {journal} {\bibinfo  {journal} {JHEP}\ }\textbf {\bibinfo {volume} {11}},\
  \bibinfo {pages} {082} (\bibinfo {year} {2009})},\ \Eprint
  {http://arxiv.org/abs/0904.0663} {arXiv:0904.0663 [hep-th]} \BibitemShut
  {NoStop}%
\bibitem [{\citenamefont {Brandhuber}\ \emph {et~al.}(2008)\citenamefont
  {Brandhuber}, \citenamefont {Heslop},\ and\ \citenamefont
  {Travaglini}}]{Brandhuber:2007yx}%
  \BibitemOpen
  \bibfield  {author} {\bibinfo {author} {\bibfnamefont {A.}~\bibnamefont
  {Brandhuber}}, \bibinfo {author} {\bibfnamefont {P.}~\bibnamefont {Heslop}},
  \ and\ \bibinfo {author} {\bibfnamefont {G.}~\bibnamefont {Travaglini}},\
  }\href {\doibase 10.1016/j.nuclphysb.2007.11.002} {\bibfield  {journal}
  {\bibinfo  {journal} {Nucl. Phys. B}\ }\textbf {\bibinfo {volume} {794}},\
  \bibinfo {pages} {231} (\bibinfo {year} {2008})},\ \Eprint
  {http://arxiv.org/abs/0707.1153} {arXiv:0707.1153 [hep-th]} \BibitemShut
  {NoStop}%
\bibitem [{\citenamefont {Drummond}\ \emph
  {et~al.}(2008{\natexlab{a}})\citenamefont {Drummond}, \citenamefont
  {Korchemsky},\ and\ \citenamefont {Sokatchev}}]{Drummond:2007aua}%
  \BibitemOpen
  \bibfield  {author} {\bibinfo {author} {\bibfnamefont {J.}~\bibnamefont
  {Drummond}}, \bibinfo {author} {\bibfnamefont {G.}~\bibnamefont
  {Korchemsky}}, \ and\ \bibinfo {author} {\bibfnamefont {E.}~\bibnamefont
  {Sokatchev}},\ }\href {\doibase 10.1016/j.nuclphysb.2007.11.041} {\bibfield
  {journal} {\bibinfo  {journal} {Nucl. Phys. B}\ }\textbf {\bibinfo {volume}
  {795}},\ \bibinfo {pages} {385} (\bibinfo {year} {2008}{\natexlab{a}})},\
  \Eprint {http://arxiv.org/abs/0707.0243} {arXiv:0707.0243 [hep-th]}
  \BibitemShut {NoStop}%
\bibitem [{\citenamefont {Drummond}\ \emph
  {et~al.}(2008{\natexlab{b}})\citenamefont {Drummond}, \citenamefont {Henn},
  \citenamefont {Korchemsky},\ and\ \citenamefont
  {Sokatchev}}]{Drummond:2007cf}%
  \BibitemOpen
  \bibfield  {author} {\bibinfo {author} {\bibfnamefont {J.}~\bibnamefont
  {Drummond}}, \bibinfo {author} {\bibfnamefont {J.}~\bibnamefont {Henn}},
  \bibinfo {author} {\bibfnamefont {G.}~\bibnamefont {Korchemsky}}, \ and\
  \bibinfo {author} {\bibfnamefont {E.}~\bibnamefont {Sokatchev}},\ }\href
  {\doibase 10.1016/j.nuclphysb.2007.11.007} {\bibfield  {journal} {\bibinfo
  {journal} {Nucl. Phys. B}\ }\textbf {\bibinfo {volume} {795}},\ \bibinfo
  {pages} {52} (\bibinfo {year} {2008}{\natexlab{b}})},\ \Eprint
  {http://arxiv.org/abs/0709.2368} {arXiv:0709.2368 [hep-th]} \BibitemShut
  {NoStop}%
\bibitem [{\citenamefont {Drummond}\ \emph
  {et~al.}(2008{\natexlab{c}})\citenamefont {Drummond}, \citenamefont {Henn},
  \citenamefont {Korchemsky},\ and\ \citenamefont
  {Sokatchev}}]{Drummond:2007bm}%
  \BibitemOpen
  \bibfield  {author} {\bibinfo {author} {\bibfnamefont {J.}~\bibnamefont
  {Drummond}}, \bibinfo {author} {\bibfnamefont {J.}~\bibnamefont {Henn}},
  \bibinfo {author} {\bibfnamefont {G.}~\bibnamefont {Korchemsky}}, \ and\
  \bibinfo {author} {\bibfnamefont {E.}~\bibnamefont {Sokatchev}},\ }\href
  {\doibase 10.1016/j.physletb.2008.03.032} {\bibfield  {journal} {\bibinfo
  {journal} {Phys. Lett. B}\ }\textbf {\bibinfo {volume} {662}},\ \bibinfo
  {pages} {456} (\bibinfo {year} {2008}{\natexlab{c}})},\ \Eprint
  {http://arxiv.org/abs/0712.4138} {arXiv:0712.4138 [hep-th]} \BibitemShut
  {NoStop}%
\bibitem [{\citenamefont {Drummond}\ \emph {et~al.}(2009)\citenamefont
  {Drummond}, \citenamefont {Henn}, \citenamefont {Korchemsky},\ and\
  \citenamefont {Sokatchev}}]{Drummond:2008aq}%
  \BibitemOpen
  \bibfield  {author} {\bibinfo {author} {\bibfnamefont {J.}~\bibnamefont
  {Drummond}}, \bibinfo {author} {\bibfnamefont {J.}~\bibnamefont {Henn}},
  \bibinfo {author} {\bibfnamefont {G.}~\bibnamefont {Korchemsky}}, \ and\
  \bibinfo {author} {\bibfnamefont {E.}~\bibnamefont {Sokatchev}},\ }\href
  {\doibase 10.1016/j.nuclphysb.2009.02.015} {\bibfield  {journal} {\bibinfo
  {journal} {Nucl. Phys. B}\ }\textbf {\bibinfo {volume} {815}},\ \bibinfo
  {pages} {142} (\bibinfo {year} {2009})},\ \Eprint
  {http://arxiv.org/abs/0803.1466} {arXiv:0803.1466 [hep-th]} \BibitemShut
  {NoStop}%
\bibitem [{\citenamefont {Bern}\ \emph {et~al.}(2008)\citenamefont {Bern},
  \citenamefont {Dixon}, \citenamefont {Kosower}, \citenamefont {Roiban},
  \citenamefont {Spradlin}, \citenamefont {Vergu},\ and\ \citenamefont
  {Volovich}}]{Bern:2008ap}%
  \BibitemOpen
  \bibfield  {author} {\bibinfo {author} {\bibfnamefont {Z.}~\bibnamefont
  {Bern}}, \bibinfo {author} {\bibfnamefont {L.}~\bibnamefont {Dixon}},
  \bibinfo {author} {\bibfnamefont {D.}~\bibnamefont {Kosower}}, \bibinfo
  {author} {\bibfnamefont {R.}~\bibnamefont {Roiban}}, \bibinfo {author}
  {\bibfnamefont {M.}~\bibnamefont {Spradlin}}, \bibinfo {author}
  {\bibfnamefont {C.}~\bibnamefont {Vergu}}, \ and\ \bibinfo {author}
  {\bibfnamefont {A.}~\bibnamefont {Volovich}},\ }\href {\doibase
  10.1103/PhysRevD.78.045007} {\bibfield  {journal} {\bibinfo  {journal} {Phys.
  Rev. D}\ }\textbf {\bibinfo {volume} {78}},\ \bibinfo {pages} {045007}
  (\bibinfo {year} {2008})},\ \Eprint {http://arxiv.org/abs/0803.1465}
  {arXiv:0803.1465 [hep-th]} \BibitemShut {NoStop}%
\bibitem [{\citenamefont {Caron-Huot}(2011{\natexlab{b}})}]{CaronHuot:2010ek}%
  \BibitemOpen
  \bibfield  {author} {\bibinfo {author} {\bibfnamefont {S.}~\bibnamefont
  {Caron-Huot}},\ }\href {\doibase 10.1007/JHEP07(2011)058} {\bibfield
  {journal} {\bibinfo  {journal} {JHEP}\ }\textbf {\bibinfo {volume} {07}},\
  \bibinfo {pages} {058} (\bibinfo {year} {2011}{\natexlab{b}})},\ \Eprint
  {http://arxiv.org/abs/1010.1167} {arXiv:1010.1167 [hep-th]} \BibitemShut
  {NoStop}%
\bibitem [{\citenamefont {Mason}\ and\ \citenamefont
  {Skinner}(2010)}]{Mason:2010yk}%
  \BibitemOpen
  \bibfield  {author} {\bibinfo {author} {\bibfnamefont {L.}~\bibnamefont
  {Mason}}\ and\ \bibinfo {author} {\bibfnamefont {D.}~\bibnamefont
  {Skinner}},\ }\href {\doibase 10.1007/JHEP12(2010)018} {\bibfield  {journal}
  {\bibinfo  {journal} {JHEP}\ }\textbf {\bibinfo {volume} {12}},\ \bibinfo
  {pages} {018} (\bibinfo {year} {2010})},\ \Eprint
  {http://arxiv.org/abs/1009.2225} {arXiv:1009.2225 [hep-th]} \BibitemShut
  {NoStop}%
\bibitem [{\citenamefont {Beisert}\ \emph {et~al.}(2012)\citenamefont {Beisert}
  \emph {et~al.}}]{Beisert:2010jr}%
  \BibitemOpen
  \bibfield  {author} {\bibinfo {author} {\bibfnamefont {N.}~\bibnamefont
  {Beisert}} \emph {et~al.},\ }\href {\doibase 10.1007/s11005-011-0529-2}
  {\bibfield  {journal} {\bibinfo  {journal} {Lett. Math. Phys.}\ }\textbf
  {\bibinfo {volume} {99}},\ \bibinfo {pages} {3} (\bibinfo {year} {2012})},\
  \Eprint {http://arxiv.org/abs/1012.3982} {arXiv:1012.3982 [hep-th]}
  \BibitemShut {NoStop}%
\bibitem [{\citenamefont {Alday}\ \emph {et~al.}(2011)\citenamefont {Alday},
  \citenamefont {Gaiotto}, \citenamefont {Maldacena}, \citenamefont {Sever},\
  and\ \citenamefont {Vieira}}]{Alday:2010ku}%
  \BibitemOpen
  \bibfield  {author} {\bibinfo {author} {\bibfnamefont {L.~F.}\ \bibnamefont
  {Alday}}, \bibinfo {author} {\bibfnamefont {D.}~\bibnamefont {Gaiotto}},
  \bibinfo {author} {\bibfnamefont {J.}~\bibnamefont {Maldacena}}, \bibinfo
  {author} {\bibfnamefont {A.}~\bibnamefont {Sever}}, \ and\ \bibinfo {author}
  {\bibfnamefont {P.}~\bibnamefont {Vieira}},\ }\href {\doibase
  10.1007/JHEP04(2011)088} {\bibfield  {journal} {\bibinfo  {journal} {JHEP}\
  }\textbf {\bibinfo {volume} {04}},\ \bibinfo {pages} {088} (\bibinfo {year}
  {2011})},\ \Eprint {http://arxiv.org/abs/1006.2788} {arXiv:1006.2788
  [hep-th]} \BibitemShut {NoStop}%
\bibitem [{\citenamefont {Basso}\ \emph {et~al.}(2013)\citenamefont {Basso},
  \citenamefont {Sever},\ and\ \citenamefont {Vieira}}]{Basso:2013vsa}%
  \BibitemOpen
  \bibfield  {author} {\bibinfo {author} {\bibfnamefont {B.}~\bibnamefont
  {Basso}}, \bibinfo {author} {\bibfnamefont {A.}~\bibnamefont {Sever}}, \ and\
  \bibinfo {author} {\bibfnamefont {P.}~\bibnamefont {Vieira}},\ }\href
  {\doibase 10.1103/PhysRevLett.111.091602} {\bibfield  {journal} {\bibinfo
  {journal} {Phys. Rev. Lett.}\ }\textbf {\bibinfo {volume} {111}},\ \bibinfo
  {pages} {091602} (\bibinfo {year} {2013})},\ \Eprint
  {http://arxiv.org/abs/1303.1396} {arXiv:1303.1396 [hep-th]} \BibitemShut
  {NoStop}%
\bibitem [{\citenamefont {Caron-Huot}\ and\ \citenamefont
  {He}(2012)}]{CaronHuot:2011kk}%
  \BibitemOpen
  \bibfield  {author} {\bibinfo {author} {\bibfnamefont {S.}~\bibnamefont
  {Caron-Huot}}\ and\ \bibinfo {author} {\bibfnamefont {S.}~\bibnamefont
  {He}},\ }\href {\doibase 10.1007/JHEP07(2012)174} {\bibfield  {journal}
  {\bibinfo  {journal} {JHEP}\ }\textbf {\bibinfo {volume} {07}},\ \bibinfo
  {pages} {174} (\bibinfo {year} {2012})},\ \Eprint
  {http://arxiv.org/abs/1112.1060} {arXiv:1112.1060 [hep-th]} \BibitemShut
  {NoStop}%
\bibitem [{Note1()}]{Note1}%
  \BibitemOpen
  \bibinfo {note} {Similar ideas have been used in~\cite {CaronHuot:2011ky}
  which motivated our investigations; they have also been explored in~\cite
  {Brandhuber:2007yx,Anastasiou:2011zk} as well.}\BibitemShut {Stop}%
\bibitem [{Note2()}]{Note2}%
  \BibitemOpen
  \bibinfo {note} {The $\protect \mathrm {d}\protect \qopname \relax o{log}$
  representation plays an important role in the study of Feynman integrals, but
  our WL-based form differs from those studied before~\cite
  {ArkaniHamed:2012nw,Herrmann:2019upk}; rather it takes a form very similar to
  those $\tau $ integrals in $\protect \bar {Q}$ calculations.}\BibitemShut
  {Stop}%
\bibitem [{\citenamefont {He}\ \emph {et~al.}(2020{\natexlab{c}})\citenamefont
  {He}, \citenamefont {Li}, \citenamefont {Tang},\ and\ \citenamefont
  {Yang}}]{He:2020uxy}%
  \BibitemOpen
  \bibfield  {author} {\bibinfo {author} {\bibfnamefont {S.}~\bibnamefont
  {He}}, \bibinfo {author} {\bibfnamefont {Z.}~\bibnamefont {Li}}, \bibinfo
  {author} {\bibfnamefont {Y.}~\bibnamefont {Tang}}, \ and\ \bibinfo {author}
  {\bibfnamefont {Q.}~\bibnamefont {Yang}},\ }\href@noop {} {\  (\bibinfo
  {year} {2020}{\natexlab{c}})},\ \Eprint {http://arxiv.org/abs/2012.13094}
  {arXiv:2012.13094 [hep-th]} \BibitemShut {NoStop}%
\bibitem [{\citenamefont {Arkani-Hamed}\ \emph {et~al.}(2012)\citenamefont
  {Arkani-Hamed}, \citenamefont {Bourjaily}, \citenamefont {Cachazo},\ and\
  \citenamefont {Trnka}}]{ArkaniHamed:2010gh}%
  \BibitemOpen
  \bibfield  {author} {\bibinfo {author} {\bibfnamefont {N.}~\bibnamefont
  {Arkani-Hamed}}, \bibinfo {author} {\bibfnamefont {J.~L.}\ \bibnamefont
  {Bourjaily}}, \bibinfo {author} {\bibfnamefont {F.}~\bibnamefont {Cachazo}},
  \ and\ \bibinfo {author} {\bibfnamefont {J.}~\bibnamefont {Trnka}},\ }\href
  {\doibase 10.1007/JHEP06(2012)125} {\bibfield  {journal} {\bibinfo  {journal}
  {JHEP}\ }\textbf {\bibinfo {volume} {06}},\ \bibinfo {pages} {125} (\bibinfo
  {year} {2012})},\ \Eprint {http://arxiv.org/abs/1012.6032} {arXiv:1012.6032
  [hep-th]} \BibitemShut {NoStop}%
\bibitem [{\citenamefont {Hodges}(2013)}]{Hodges:2009hk}%
  \BibitemOpen
  \bibfield  {author} {\bibinfo {author} {\bibfnamefont {A.}~\bibnamefont
  {Hodges}},\ }\href {\doibase 10.1007/JHEP05(2013)135} {\bibfield  {journal}
  {\bibinfo  {journal} {JHEP}\ }\textbf {\bibinfo {volume} {05}},\ \bibinfo
  {pages} {135} (\bibinfo {year} {2013})},\ \Eprint
  {http://arxiv.org/abs/0905.1473} {arXiv:0905.1473 [hep-th]} \BibitemShut
  {NoStop}%
\bibitem [{\citenamefont {Drummond}\ \emph {et~al.}(2007)\citenamefont
  {Drummond}, \citenamefont {Henn}, \citenamefont {Smirnov},\ and\
  \citenamefont {Sokatchev}}]{Drummond:2006rz}%
  \BibitemOpen
  \bibfield  {author} {\bibinfo {author} {\bibfnamefont {J.~M.}\ \bibnamefont
  {Drummond}}, \bibinfo {author} {\bibfnamefont {J.}~\bibnamefont {Henn}},
  \bibinfo {author} {\bibfnamefont {V.~A.}\ \bibnamefont {Smirnov}}, \ and\
  \bibinfo {author} {\bibfnamefont {E.}~\bibnamefont {Sokatchev}},\ }\href
  {\doibase 10.1088/1126-6708/2007/01/064} {\bibfield  {journal} {\bibinfo
  {journal} {JHEP}\ }\textbf {\bibinfo {volume} {01}},\ \bibinfo {pages} {064}
  (\bibinfo {year} {2007})},\ \Eprint {http://arxiv.org/abs/hep-th/0607160}
  {arXiv:hep-th/0607160 [hep-th]} \BibitemShut {NoStop}%
\bibitem [{\citenamefont {Drummond}\ \emph {et~al.}(2010)\citenamefont
  {Drummond}, \citenamefont {Henn}, \citenamefont {Korchemsky},\ and\
  \citenamefont {Sokatchev}}]{Drummond:2008vq}%
  \BibitemOpen
  \bibfield  {author} {\bibinfo {author} {\bibfnamefont {J.~M.}\ \bibnamefont
  {Drummond}}, \bibinfo {author} {\bibfnamefont {J.}~\bibnamefont {Henn}},
  \bibinfo {author} {\bibfnamefont {G.~P.}\ \bibnamefont {Korchemsky}}, \ and\
  \bibinfo {author} {\bibfnamefont {E.}~\bibnamefont {Sokatchev}},\ }\href
  {\doibase 10.1016/j.nuclphysb.2009.11.022} {\bibfield  {journal} {\bibinfo
  {journal} {Nucl. Phys.}\ }\textbf {\bibinfo {volume} {B828}},\ \bibinfo
  {pages} {317} (\bibinfo {year} {2010})},\ \Eprint
  {http://arxiv.org/abs/0807.1095} {arXiv:0807.1095 [hep-th]} \BibitemShut
  {NoStop}%
\bibitem [{\citenamefont {Korchemsky}\ and\ \citenamefont
  {Sokatchev}(2010)}]{Korchemsky:2010ut}%
  \BibitemOpen
  \bibfield  {author} {\bibinfo {author} {\bibfnamefont {G.~P.}\ \bibnamefont
  {Korchemsky}}\ and\ \bibinfo {author} {\bibfnamefont {E.}~\bibnamefont
  {Sokatchev}},\ }\href {\doibase 10.1016/j.nuclphysb.2010.05.022} {\bibfield
  {journal} {\bibinfo  {journal} {Nucl. Phys.}\ }\textbf {\bibinfo {volume}
  {B839}},\ \bibinfo {pages} {377} (\bibinfo {year} {2010})},\ \Eprint
  {http://arxiv.org/abs/1002.4625} {arXiv:1002.4625 [hep-th]} \BibitemShut
  {NoStop}%
\bibitem [{Note3()}]{Note3}%
  \BibitemOpen
  \bibinfo {note} {Note that no diagram with less insertion is possible, which
  explains why the component vanishes at tree and one-loop level, and it is
  finite at two-loop level.}\BibitemShut {Stop}%
\bibitem [{\citenamefont {Dixon}\ \emph {et~al.}(2012)\citenamefont {Dixon},
  \citenamefont {Drummond},\ and\ \citenamefont {Henn}}]{Dixon:2011nj}%
  \BibitemOpen
  \bibfield  {author} {\bibinfo {author} {\bibfnamefont {L.~J.}\ \bibnamefont
  {Dixon}}, \bibinfo {author} {\bibfnamefont {J.~M.}\ \bibnamefont {Drummond}},
  \ and\ \bibinfo {author} {\bibfnamefont {J.~M.}\ \bibnamefont {Henn}},\
  }\href {\doibase 10.1007/JHEP01(2012)024} {\bibfield  {journal} {\bibinfo
  {journal} {JHEP}\ }\textbf {\bibinfo {volume} {01}},\ \bibinfo {pages} {024}
  (\bibinfo {year} {2012})},\ \Eprint {http://arxiv.org/abs/1111.1704}
  {arXiv:1111.1704 [hep-th]} \BibitemShut {NoStop}%
\bibitem [{Note4()}]{Note4}%
  \BibitemOpen
  \bibinfo {note} {For two-dimensional kinematics, $n=8$ double pentagons have
  been evaluated analytically in~\cite {Alday:2010jz}; see~\cite
  {Heslop:2010kq} and \cite {Caron-Huot:2013vda} for analytic results of
  two-loop and three-loop amplitudes in two-dimensional
  kinematics.}\BibitemShut {Stop}%
\bibitem [{\citenamefont {Bourjaily}\ \emph {et~al.}(2020)\citenamefont
  {Bourjaily}, \citenamefont {McLeod}, \citenamefont {Vergu}, \citenamefont
  {Volk}, \citenamefont {Von~Hippel},\ and\ \citenamefont
  {Wilhelm}}]{Bourjaily:2019igt}%
  \BibitemOpen
  \bibfield  {author} {\bibinfo {author} {\bibfnamefont {J.~L.}\ \bibnamefont
  {Bourjaily}}, \bibinfo {author} {\bibfnamefont {A.~J.}\ \bibnamefont
  {McLeod}}, \bibinfo {author} {\bibfnamefont {C.}~\bibnamefont {Vergu}},
  \bibinfo {author} {\bibfnamefont {M.}~\bibnamefont {Volk}}, \bibinfo {author}
  {\bibfnamefont {M.}~\bibnamefont {Von~Hippel}}, \ and\ \bibinfo {author}
  {\bibfnamefont {M.}~\bibnamefont {Wilhelm}},\ }\href {\doibase
  10.1007/JHEP02(2020)025} {\bibfield  {journal} {\bibinfo  {journal} {JHEP}\
  }\textbf {\bibinfo {volume} {02}},\ \bibinfo {pages} {025} (\bibinfo {year}
  {2020})},\ \Eprint {http://arxiv.org/abs/1910.14224} {arXiv:1910.14224
  [hep-th]} \BibitemShut {NoStop}%
\bibitem [{\citenamefont {Goncharov}\ \emph {et~al.}(2010)\citenamefont
  {Goncharov}, \citenamefont {Spradlin}, \citenamefont {Vergu},\ and\
  \citenamefont {Volovich}}]{Goncharov:2010jf}%
  \BibitemOpen
  \bibfield  {author} {\bibinfo {author} {\bibfnamefont {A.~B.}\ \bibnamefont
  {Goncharov}}, \bibinfo {author} {\bibfnamefont {M.}~\bibnamefont {Spradlin}},
  \bibinfo {author} {\bibfnamefont {C.}~\bibnamefont {Vergu}}, \ and\ \bibinfo
  {author} {\bibfnamefont {A.}~\bibnamefont {Volovich}},\ }\href {\doibase
  10.1103/PhysRevLett.105.151605} {\bibfield  {journal} {\bibinfo  {journal}
  {Phys. Rev. Lett.}\ }\textbf {\bibinfo {volume} {105}},\ \bibinfo {pages}
  {151605} (\bibinfo {year} {2010})},\ \Eprint {http://arxiv.org/abs/1006.5703}
  {arXiv:1006.5703 [hep-th]} \BibitemShut {NoStop}%
\bibitem [{\citenamefont {Duhr}\ \emph {et~al.}(2012)\citenamefont {Duhr},
  \citenamefont {Gangl},\ and\ \citenamefont {Rhodes}}]{Duhr:2011zq}%
  \BibitemOpen
  \bibfield  {author} {\bibinfo {author} {\bibfnamefont {C.}~\bibnamefont
  {Duhr}}, \bibinfo {author} {\bibfnamefont {H.}~\bibnamefont {Gangl}}, \ and\
  \bibinfo {author} {\bibfnamefont {J.~R.}\ \bibnamefont {Rhodes}},\ }\href
  {\doibase 10.1007/JHEP10(2012)075} {\bibfield  {journal} {\bibinfo  {journal}
  {JHEP}\ }\textbf {\bibinfo {volume} {10}},\ \bibinfo {pages} {075} (\bibinfo
  {year} {2012})},\ \Eprint {http://arxiv.org/abs/1110.0458} {arXiv:1110.0458
  [math-ph]} \BibitemShut {NoStop}%
\bibitem [{\citenamefont {Chicherin}\ \emph {et~al.}(2013)\citenamefont
  {Chicherin}, \citenamefont {Derkachov},\ and\ \citenamefont
  {Isaev}}]{Chicherin:2012yn}%
  \BibitemOpen
  \bibfield  {author} {\bibinfo {author} {\bibfnamefont {D.}~\bibnamefont
  {Chicherin}}, \bibinfo {author} {\bibfnamefont {S.}~\bibnamefont
  {Derkachov}}, \ and\ \bibinfo {author} {\bibfnamefont {A.}~\bibnamefont
  {Isaev}},\ }\href {\doibase 10.1007/JHEP04(2013)020} {\bibfield  {journal}
  {\bibinfo  {journal} {JHEP}\ }\textbf {\bibinfo {volume} {04}},\ \bibinfo
  {pages} {020} (\bibinfo {year} {2013})},\ \Eprint
  {http://arxiv.org/abs/1206.4150} {arXiv:1206.4150 [math-ph]} \BibitemShut
  {NoStop}%
\bibitem [{\citenamefont {Bern}\ \emph {et~al.}(1993)\citenamefont {Bern},
  \citenamefont {Dixon},\ and\ \citenamefont {Kosower}}]{Bern:1992em}%
  \BibitemOpen
  \bibfield  {author} {\bibinfo {author} {\bibfnamefont {Z.}~\bibnamefont
  {Bern}}, \bibinfo {author} {\bibfnamefont {L.~J.}\ \bibnamefont {Dixon}}, \
  and\ \bibinfo {author} {\bibfnamefont {D.~A.}\ \bibnamefont {Kosower}},\
  }\href {\doibase 10.1016/0370-2693(93)90400-C} {\bibfield  {journal}
  {\bibinfo  {journal} {Phys. Lett. B}\ }\textbf {\bibinfo {volume} {302}},\
  \bibinfo {pages} {299} (\bibinfo {year} {1993})},\ \bibinfo {note} {[Erratum:
  Phys.Lett.B 318, 649 (1993)]},\ \Eprint {http://arxiv.org/abs/hep-ph/9212308}
  {arXiv:hep-ph/9212308} \BibitemShut {NoStop}%
\bibitem [{\citenamefont {Bern}\ \emph {et~al.}(1994)\citenamefont {Bern},
  \citenamefont {Dixon},\ and\ \citenamefont {Kosower}}]{Bern:1993kr}%
  \BibitemOpen
  \bibfield  {author} {\bibinfo {author} {\bibfnamefont {Z.}~\bibnamefont
  {Bern}}, \bibinfo {author} {\bibfnamefont {L.~J.}\ \bibnamefont {Dixon}}, \
  and\ \bibinfo {author} {\bibfnamefont {D.~A.}\ \bibnamefont {Kosower}},\
  }\href {\doibase 10.1016/0550-3213(94)90398-0} {\bibfield  {journal}
  {\bibinfo  {journal} {Nucl. Phys. B}\ }\textbf {\bibinfo {volume} {412}},\
  \bibinfo {pages} {751} (\bibinfo {year} {1994})},\ \Eprint
  {http://arxiv.org/abs/hep-ph/9306240} {arXiv:hep-ph/9306240} \BibitemShut
  {NoStop}%
\bibitem [{\citenamefont {Arkani-Hamed}\ and\ \citenamefont
  {Yuan}(2017)}]{Arkani-Hamed:2017ahv}%
  \BibitemOpen
  \bibfield  {author} {\bibinfo {author} {\bibfnamefont {N.}~\bibnamefont
  {Arkani-Hamed}}\ and\ \bibinfo {author} {\bibfnamefont {E.~Y.}\ \bibnamefont
  {Yuan}},\ }\href@noop {} {\  (\bibinfo {year} {2017})},\ \Eprint
  {http://arxiv.org/abs/1712.09991} {arXiv:1712.09991 [hep-th]} \BibitemShut
  {NoStop}%
\bibitem [{Note5()}]{Note5}%
  \BibitemOpen
  \bibinfo {note} {Throughout the letter, any relabelling acts on external
  legs/momentum twistors, rather than on dual points. We denote the $15$
  residues of the hexagon ($2$ forms) by the two propagators that are {\protect
  \it not} cut in each quadruple cut, {\protect \it e.g.} $[x,y]$ denotes the
  residue with $x_k, x_{k{+}1}, x_l, x_{l{+}1}$ cut; by Global Residue Theorem
  (GRT) only $5$ of them are independent, which we choose to be $[x,y], [x,
  x_k], \protect \cdots , [x, x_{l{+}1}]$.}\BibitemShut {Stop}%
\bibitem [{Note6()}]{Note6}%
  \BibitemOpen
  \bibinfo {note} {We still need to perform ``rationalization'' and encounter
  spurious square roots of kinematics from $w$ , but they all nicely cancel in
  the final answer as expected.}\BibitemShut {Stop}%
\bibitem [{Note7()}]{Note7}%
  \BibitemOpen
  \bibinfo {note} {Note that when converting ${\protect \rm d}\protect \qopname
  \relax o{log}$'s of $\tau '$ into those of $z(\tau ')$, we omitted
  ``constants'' that are independent of $\tau '$; it is crucial to also drop
  the same constants for ${\protect \rm d}\protect \qopname \relax o{log}$'s in
  the rational part, since one can only ignore such constants for the entire,
  integrable symbol.}\BibitemShut {Stop}%
\bibitem [{\citenamefont {Dennen}\ \emph {et~al.}(2016)\citenamefont {Dennen},
  \citenamefont {Spradlin},\ and\ \citenamefont {Volovich}}]{Dennen:2015bet}%
  \BibitemOpen
  \bibfield  {author} {\bibinfo {author} {\bibfnamefont {T.}~\bibnamefont
  {Dennen}}, \bibinfo {author} {\bibfnamefont {M.}~\bibnamefont {Spradlin}}, \
  and\ \bibinfo {author} {\bibfnamefont {A.}~\bibnamefont {Volovich}},\ }\href
  {\doibase 10.1007/JHEP03(2016)069} {\bibfield  {journal} {\bibinfo  {journal}
  {JHEP}\ }\textbf {\bibinfo {volume} {03}},\ \bibinfo {pages} {069} (\bibinfo
  {year} {2016})},\ \Eprint {http://arxiv.org/abs/1512.07909} {arXiv:1512.07909
  [hep-th]} \BibitemShut {NoStop}%
\bibitem [{\citenamefont {Bourjaily}\ and\ \citenamefont
  {Trnka}(2015)}]{Bourjaily:2015jna}%
  \BibitemOpen
  \bibfield  {author} {\bibinfo {author} {\bibfnamefont {J.~L.}\ \bibnamefont
  {Bourjaily}}\ and\ \bibinfo {author} {\bibfnamefont {J.}~\bibnamefont
  {Trnka}},\ }\href {\doibase 10.1007/JHEP08(2015)119} {\bibfield  {journal}
  {\bibinfo  {journal} {JHEP}\ }\textbf {\bibinfo {volume} {08}},\ \bibinfo
  {pages} {119} (\bibinfo {year} {2015})},\ \Eprint
  {http://arxiv.org/abs/1505.05886} {arXiv:1505.05886 [hep-th]} \BibitemShut
  {NoStop}%
\bibitem [{\citenamefont {Drummond}\ \emph {et~al.}(2011)\citenamefont
  {Drummond}, \citenamefont {Henn},\ and\ \citenamefont
  {Trnka}}]{Drummond:2010cz}%
  \BibitemOpen
  \bibfield  {author} {\bibinfo {author} {\bibfnamefont {J.~M.}\ \bibnamefont
  {Drummond}}, \bibinfo {author} {\bibfnamefont {J.~M.}\ \bibnamefont {Henn}},
  \ and\ \bibinfo {author} {\bibfnamefont {J.}~\bibnamefont {Trnka}},\ }\href
  {\doibase 10.1007/JHEP04(2011)083} {\bibfield  {journal} {\bibinfo  {journal}
  {JHEP}\ }\textbf {\bibinfo {volume} {04}},\ \bibinfo {pages} {083} (\bibinfo
  {year} {2011})},\ \Eprint {http://arxiv.org/abs/1010.3679} {arXiv:1010.3679
  [hep-th]} \BibitemShut {NoStop}%
\bibitem [{\citenamefont {Henn}(2013)}]{Henn:2013pwa}%
  \BibitemOpen
  \bibfield  {author} {\bibinfo {author} {\bibfnamefont {J.~M.}\ \bibnamefont
  {Henn}},\ }\href {\doibase 10.1103/PhysRevLett.110.251601} {\bibfield
  {journal} {\bibinfo  {journal} {Phys. Rev. Lett.}\ }\textbf {\bibinfo
  {volume} {110}},\ \bibinfo {pages} {251601} (\bibinfo {year} {2013})},\
  \Eprint {http://arxiv.org/abs/1304.1806} {arXiv:1304.1806 [hep-th]}
  \BibitemShut {NoStop}%
\bibitem [{\citenamefont {Anastasiou}\ and\ \citenamefont
  {Banfi}(2011)}]{Anastasiou:2011zk}%
  \BibitemOpen
  \bibfield  {author} {\bibinfo {author} {\bibfnamefont {C.}~\bibnamefont
  {Anastasiou}}\ and\ \bibinfo {author} {\bibfnamefont {A.}~\bibnamefont
  {Banfi}},\ }\href {\doibase 10.1007/JHEP02(2011)064} {\bibfield  {journal}
  {\bibinfo  {journal} {JHEP}\ }\textbf {\bibinfo {volume} {02}},\ \bibinfo
  {pages} {064} (\bibinfo {year} {2011})},\ \Eprint
  {http://arxiv.org/abs/1101.4118} {arXiv:1101.4118 [hep-th]} \BibitemShut
  {NoStop}%
\bibitem [{\citenamefont {Alday}(2011)}]{Alday:2010jz}%
  \BibitemOpen
  \bibfield  {author} {\bibinfo {author} {\bibfnamefont {L.~F.}\ \bibnamefont
  {Alday}},\ }\href {\doibase 10.1007/JHEP07(2011)080} {\bibfield  {journal}
  {\bibinfo  {journal} {JHEP}\ }\textbf {\bibinfo {volume} {07}},\ \bibinfo
  {pages} {080} (\bibinfo {year} {2011})},\ \Eprint
  {http://arxiv.org/abs/1009.1110} {arXiv:1009.1110 [hep-th]} \BibitemShut
  {NoStop}%
\bibitem [{\citenamefont {Heslop}\ and\ \citenamefont
  {Khoze}(2010)}]{Heslop:2010kq}%
  \BibitemOpen
  \bibfield  {author} {\bibinfo {author} {\bibfnamefont {P.}~\bibnamefont
  {Heslop}}\ and\ \bibinfo {author} {\bibfnamefont {V.~V.}\ \bibnamefont
  {Khoze}},\ }\href {\doibase 10.1007/JHEP11(2010)035} {\bibfield  {journal}
  {\bibinfo  {journal} {JHEP}\ }\textbf {\bibinfo {volume} {11}},\ \bibinfo
  {pages} {035} (\bibinfo {year} {2010})},\ \Eprint
  {http://arxiv.org/abs/1007.1805} {arXiv:1007.1805 [hep-th]} \BibitemShut
  {NoStop}%
\bibitem [{\citenamefont {Caron-Huot}\ and\ \citenamefont
  {He}(2013)}]{Caron-Huot:2013vda}%
  \BibitemOpen
  \bibfield  {author} {\bibinfo {author} {\bibfnamefont {S.}~\bibnamefont
  {Caron-Huot}}\ and\ \bibinfo {author} {\bibfnamefont {S.}~\bibnamefont
  {He}},\ }\href {\doibase 10.1007/JHEP08(2013)101} {\bibfield  {journal}
  {\bibinfo  {journal} {JHEP}\ }\textbf {\bibinfo {volume} {08}},\ \bibinfo
  {pages} {101} (\bibinfo {year} {2013})},\ \Eprint
  {http://arxiv.org/abs/1305.2781} {arXiv:1305.2781 [hep-th]} \BibitemShut
  {NoStop}%
\end{thebibliography}%


\widetext
\clearpage
\begin{center}
\textbf{\large Appendix}
\end{center}
\setcounter{equation}{0}
\setcounter{figure}{0}
\setcounter{table}{0}
\makeatletter
\renewcommand{\theequation}{A\arabic{equation}}
\renewcommand{\thefigure}{A\arabic{figure}}
\renewcommand{\bibnumfmt}[1]{[A#1]}
\renewcommand{\citenumfont}[1]{A#1}

Here we record the symbol of the two weight-$3$ functions for $I_{\rm dp}(1,4,7,10)$ with $n=12$: $R^{\bar 1}_{3 4}$ and $M^{1 7 \,10}_{34}$. The result is DCI and depends on $11$ multiplicatively-independent cross ratios:
\begin{align*}
u_{1}&=\frac{x_{1,4}^2 x_{2,7}^2}{x_{2,4}^2 x_{1,7}^2},u_{2}=\frac{x_{1,4}^2 x_{2,8}^2}{x_{2,4}^2 x_{1,8}^2},u_{3}=\frac{x_{1,4}^2 x_{2,10}^2}{x_{2,4}^2 x_{1,10}^2},u_{4}=\frac{x_{1,4}^2 x_{2,11}^2}{x_{2,4}^2 x_{1,11}^2},u_{5}=\frac{x_{2,7}^2 x_{4,8}^2}{x_{2,8}^2 x_{4,7}^2},u_{6}=\frac{x_{4,8}^2 x_{7,10}^2}{x_{4,7}^2 x_{8,10}^2},\\
u_{7}&=\frac{x_{4,8}^2 x_{7,11}^2}{x_{4,7}^2 x_{8,11}^2},u_{8}=\frac{x_{2,10}^2 x_{7,11}^2}{x_{2,11}^2 x_{7,10}^2},u_{9}=\frac{x_{2,7}^2 x_{4,10}^2}{x_{2,10}^2 x_{4,7}^2},u_{10}=\frac{x_{2,4}^2 x_{7,10}^2}{x_{2,7}^2 x_{4,10}^2},u_{11}=\frac{x_{2,11}^2 x_{4,7}^2}{x_{2,7}^2 x_{4,11}^2}
\end{align*}

We first write the symbol of $M$, which consists of an algebraic part and a rational part $ \mathcal{S}(M_{34}^{1 7\,10})=S_{\rm{alg}.}+S_{\rm{rat}.}$. The algebraic part is obtained by extracting the weight-$3$ coefficient of $\langle 1 3 4 7\rangle$ of algebraic words (see \eqref{algwords}):
{\small \begin{equation}
    S_{\rm{alg}.}=\mathcal{S}(F(x_4,x_2,x_7,x_{10}))\otimes\left(\frac{\frac{\langle x_2 x_4 \rangle\langle x_{10} 17 \rangle }{\langle x_2 x_{10} \rangle \langle x_{4}17 \rangle}-z(x_4,x_2,x_7,x_{10})}{\frac{\langle x_2 x_4 \rangle\langle x_{10} 17 \rangle }{\langle x_2 x_{10} \rangle \langle x_{4}17 \rangle}-\bar{z}(x_4,x_2,x_7,x_{10})}\right)-(9\leftrightarrow 11)-(6\leftrightarrow 8)-(2\leftrightarrow 12)\,,
\end{equation}}
where as usual each relabelling applies to all previous terms; the rational part reads
{\small
\begin{align}
S_{\rm{rat}.}&=\mathcal{S}(f_1)\otimes \frac{\langle 127\,12\rangle \langle 1(34)(67)(9\,10)\rangle}{\langle 179\,10\rangle \langle 1(2\,12)(34)(67)\rangle}-(6 \leftrightarrow 8)+\mathcal{S}(f_2)\otimes \frac{\langle 134\,12\rangle \langle 127\,12\rangle \langle 3478\rangle \langle 179\,10\rangle}{\langle 1(2\,12)(34)(78)\rangle \langle 7(1\,12)(34)(9\,10)\rangle}\\\nonumber
&+\mathcal{S}(f_3)\otimes \frac{\langle 1234\rangle \langle 7(1\,12)(34)(9\,10)\rangle}{\langle 134\,12\rangle \langle 7(12)(34)(9\,10)\rangle}+\mathcal{S}(I_{\rm p}\left(u_1,\frac{u_1 u_6}{u_2 u_5},\frac{u_2}{u_1}\right))\otimes \frac{\langle 1234\rangle \langle 127\,12\rangle \langle 179\,10\rangle \langle 7(34)(68)(9\,10)\rangle}{\langle 7(12)(34)(9\,10)\rangle \langle \bar 7 \cap (79\,10)\cap \bar 1\cap (134)))}-(9 \leftrightarrow 11)\\\nonumber
&+\mathcal{S}(f_4)\otimes \frac{\langle 7(34)(68)(10\,11)\rangle \langle 7(12)(34)(9\,10)\rangle}{\langle 7(34)(68)(9\,10)\rangle \langle 7(12)(34)(10\,11)\rangle}+\mathcal{S}(f_{5})\otimes \frac{\langle 3478\rangle \langle 1(2\,12)(34)(67)\rangle}{\langle 3467\rangle \langle 1(2\,12)(34)(78)\rangle}
\end{align}}
where we represent the first two-entries as symbol of dilogarithm functions
\begin{equation*}
\begin{split}
f_1&=-\operatorname{Li}_2(1{-}u_3)+\operatorname{Li}_2(1{-}u_1)+\operatorname{Li}_2\left(1{-}\frac{u_3}{u_1}\right)-\frac{1}{2} \log (u_1) \log \left(\frac{u_3}{u_1}\right)-\frac{1}{2} \log (u_1) \log (u_{10} u_9)+\frac{1}{2} \log (u_{10}) \log (u_3)\\
f_2&=\frac{1}{2} \log \left(\frac{u_2}{u_1}\right) \log (u_{10} u_3 u_9)-\frac{1}{2} \log (u_6) \log (u_2 u_3)+\frac{1}{2} \log (u_3) \log (u_5)\\
f_3&=\operatorname{Li}_2(1{-}u_6)-\operatorname{Li}_2\biggl(1{-}\frac{u_2 u_5}{u_1}\biggr)-\operatorname{Li}_2\biggl(1{-}\frac{u_1 u_6}{u_2 u_5}\biggr)+\frac{1}{2} \log \biggl(\frac{u_2 u_5}{u_1}\biggr) \log \biggl(\frac{u_1 u_6}{u_2 u_5}\biggr)+\frac{1}{2} \log (u_6) \log \left(\frac{u_9}{u_1 u_2}\right)\\
&\quad-\frac{1}{2} \log (u_5) \log (u_{10} u_9)\\
f_4&=I_{\rm p}\left(u_1,u_5,\frac{u_2}{u_1}\right)+\frac12\log (u_1)^2-\frac12\log (u_2)^2-\log (u_5) \log (u_1 u_2)\\
f_{5}&=-\frac{1}{2} \log (u_4) \log (u_1 u_{10} u_{11} u_8 u_9)+\frac{1}{2} \log (u_3) \log (u_1 u_{10})-\frac{1}{2} \log (u_1) \log (u_8)
\end{split}
\end{equation*}
and $I_{\rm p}$ is the one loop chiral pentagon $I_{\rm p}(u,v,w)=-\operatorname{Li}_2(1-u w)+\operatorname{Li}_2(1-u)-\operatorname{Li}_2(1-v w)+\operatorname{Li}_2(1-v)+\operatorname{Li}_2(1-w)+\log (u) \log (v)$.
Similarly the symbol of $R$, which has only rational letters, can be written as 
{\small  
\begin{align}
\mathcal{S}(R^{\bar 1}_{3 4})&=\biggl(\mathcal{S}(f_{6})\otimes \frac{\langle 12\,1\,3\,4\rangle \langle 12\,1\,2\,7\rangle \langle 1\,6\,7\,10\rangle}{\langle 12\,1\,7\,10\rangle \langle 1(12\,2)(3\,4)(6\,7)\rangle}-(6\leftrightarrow 8)\biggr)\\\nonumber
&+\biggl(\mathcal{S}(I_{\rm p}\left(u_{1},\frac{u_{1} u_{6}}{u_{2} u_{5}},\frac{u_{2}}{u_{1}}\right))\otimes \frac{\langle 12\,1\,7\,10\rangle \langle \bar 1\cap (1\,3\,4)\cap \bar 7 \cap (7\,9\,10)\rangle}{\langle 12\,1\,3\,4\rangle \langle 12\,1\,2\,7\rangle \langle 6\,7\,8\,10\rangle \langle 1\,7\,9\,10\rangle}-(9\leftrightarrow 11)\biggr)\\\nonumber
&+(6\leftrightarrow 9,7\leftrightarrow 10,8\leftrightarrow 11)+\mathcal{S}(I_{\rm p}\left(u_{8},\frac{u_{5}}{u_{6}},\frac{u_{6}}{u_{7}}\right))\otimes \frac{\langle 12\,1\,3\,4\rangle \langle 1\,2\,7\,10\rangle}{\langle 1\,2\,3\,4\rangle \langle 12\,1\,7\,10\rangle}
\end{align}}where on the first two lines the relabelling applies only to its previous term, while $(6\leftrightarrow 9,7\leftrightarrow 10,8\leftrightarrow 11)$ applies to all terms above; the first two entries are given by symbol of various $I_{\rm p}$'s and that of
\begin{equation*}
    f_{6}=\operatorname{Li}_2\left(1-\frac{u_{1}}{u_{3}}\right)-\operatorname{Li}_2\left(1-\frac{u_{1}}{u_{4}}\right)-\operatorname{Li}_2\left(1-\frac{1}{u_{3}}\right)+\operatorname{Li}_2\left(1-\frac{1}{u_{4}}\right)-\log (u_{1}) \log (u_{8})\,.
\end{equation*}

\end{document}